\documentclass[10pt,journal,compsoc]{IEEEtran}
\usepackage{cite}
\usepackage{graphicx}
\usepackage{amsmath}
\usepackage{algorithmic}
\usepackage{array}
\usepackage{subcaption}
\usepackage{url}
\usepackage{wrapfig}

\begin{document}

\title{Fast Multiplex Graph Association Rules for Link Prediction}

\author{Michele Coscia, \IEEEmembership{IT University of Copenhagen};
        Christian Borgelt, \IEEEmembership{University of Salzburg};\\
        and Michael Szell, \IEEEmembership{IT University of Copenhagen}\\
E-mail: mcos@itu.dk}

\IEEEtitleabstractindextext{%
\begin{abstract}
Multiplex networks allow us to study a variety of complex systems where nodes connect to each other in multiple ways, for example friend, family, and co-worker relations in social networks. Link prediction is the branch of network analysis allowing us to forecast the future status of a network: which new connections are the most likely to appear in the future? In multiplex link prediction we also ask: of which type? Because this last question is unanswerable with classical link prediction, here we investigate the use of graph association rules to inform multiplex link prediction. We derive such rules by identifying all frequent patterns in a network via multiplex graph mining, and then score each unobserved link's likelihood by finding the occurrences of each rule in the original network. Association rules add new abilities to multiplex link prediction: to predict new node arrivals, to consider higher order structures with four or more nodes, and to be memory efficient. We improve over previous work by creating a framework that is also efficient in terms of runtime, which enables an increase in prediction performance. This increase in efficiency allows us to improve a case study on a signed multiplex network, showing how graph association rules can provide valuable insights to extend social balance theory.
\end{abstract}

}

\maketitle

\IEEEdisplaynontitleabstractindextext

\IEEEpeerreviewmaketitle

\section{Introduction}
Complex networks are a powerful abstraction of interacting entities (nodes and links), well suited to study complex systems, from society, the brain, to interdependent infrastructure services. Given its analytical power, network analysis can be used to forecast the future status of a system, for instance to predict new relationships or routes. This is the well-known problem of link prediction: the task of estimating the likelihood of unobserved connections to be observed in the future \cite{liben2007link,lu2011link,zhang2018link}. A powerful technique to solve link prediction in simple networks is graph association rules as in \texttt{GERM} \cite{berlingerio2009mining,bringmann2010learning} and \texttt{MAGMA} \cite{coscia2020multiplex}, our prior work.

\begin{figure}
\centering
\includegraphics[width=.6\columnwidth]{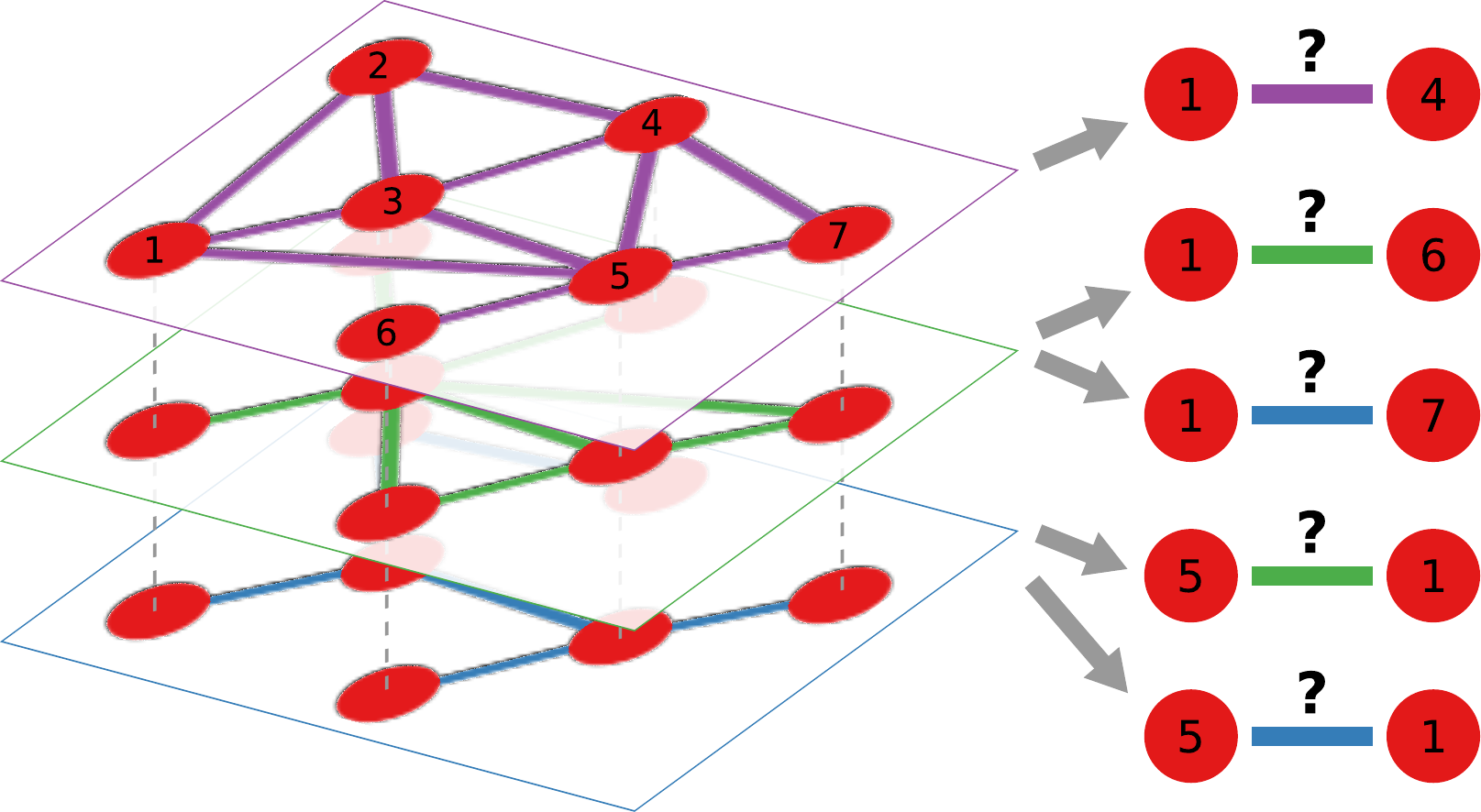}
\caption{A depiction of the multiplex link prediction problem. Will two nodes connect, and how?}
\label{fig:mllp}
\end{figure}

In multiplex networks, entities can connect in different ways \cite{krackhardt1987cognitive,roethlisberger1939management,kivela2014multilayer,boccaletti2014structure}. For example, you know people for different reasons -- friendship, work ties, economic transactions --, or you move through space with different means of transportation -- bicycle, car, train, plane \cite{berlingerio2011foundations,dickison2016multilayer}. Multiplex link prediction \cite{rossetti2011scalable,matsuno2018mell} comes with additional challenges, as Figure \ref{fig:mllp} shows. Here, we are not only interested in knowing \textit{whether} two nodes will connect: we also want to know \textit{how} they will connect. In fact, as the figure shows, nodes that are already connected can also be part of the solution space: for example, nodes $1$ and $5$ are connected by a link in the top layer, but not in the middle and bottom ones. Thus, we want to estimate the likelihood they are going to connect via edges of multiple types -- or from multiple layers, since multiplex networks are a special subtype of multilayer networks (throughout the paper we use the terms ``type'' and ``layer'' interchangeably).

There are other approaches to perform multiplex link prediction \cite{pujari2015link,jalili2017link,sharma2016efficient,hristova2016multilayer,de2017community}. Most of them share a strategy: calculate classical link prediction scores based on the topology of the network, and then combine them into a multiplex score. Approaches based on graph neural networks are also gaining popularity, the only one that can handle multiplex networks -- to the best of our knowledge -- is Mell \cite{matsuno2018mell}. These are the methods we compare with.

In this paper we extend the usage of graph association rules to perform link prediction on multiplex networks and prove their advantages over the alternatives. We need to significantly change \texttt{GERM}'s framework because graph association rules are based on frequent graph pattern mining, which can handle only a single label on the edges. \texttt{GERM} uses the attribute to indicate the link's appearance time, while here we need to use it to indicate its type.

First, we use \texttt{MoSS} \cite{borgelt2005moss} to perform multiplex graph mining, i.e. the discovery of all frequent multiplex patterns. Then, we use these patterns to build multiplex graph association rules, connecting two frequent patterns that differ by one link. Finally, we find all occurrences of a rule in the original graph and score the likelihood of the new link to appear. 

Our previous work improved over the state of the art by extending social balance theory to structures larger than a triangle, by empowering the user to predict incoming nodes in the network, and by its memory efficiency. The advanced framework we propose in this paper has all of these advantages, and in addition it improves over the prior work in three other directions:

\textbf{Time efficiency}. By embedding the rule detection in the mining step, we save considerable computational time.

\textbf{Increased predictive performance}. As a consequence of the increase in efficiency, we can lower the support threshold parameter, which implies more patterns found, more rules generated and more precise prediction scores, at the same cost of computational time.

\textbf{Framework improvements}. Edge directions are now properly handled and we have re-vamped the definition of edge lift.

Further, we extend the baseline comparisons to include state-of-the-art neural network approaches. In our experiments, we show that, using graph association rules, we can achieve higher Area Under the ROC Curve (AUC) performance over several data sets representing systems coming from different fields, from web-mediated online social interactions to neural networks.

Our framework can perform single layer link prediction via graph association rules, not just multiplex link prediction. It is open-source and freely available. The code is part of the new edition of the \texttt{Moss} software, available at \url{http://www.borgelt.net/moss.html}. The data and scripts to replicate the paper's results are available at \url{http://www.michelecoscia.com/?page_id=1857}. This package also includes our implementations of other multiplex link prediction techniques. All external dependencies required to properly run the baseline methods are available in the package or the package contains instructions on how to retrieve them. Being a neural network approach, the Mell baseline requires a working Tensorflow installation.

\section{Related Work}
In this paper we address multiplex link prediction via mining graph association rules, implying that we need to perform frequent pattern mining over a multiplex network.

\subsection{Frequent Graph Pattern Mining}\label{sec:related-fpm}
Frequent pattern mining in graphs is the search for frequent subgraph patterns \cite{chakrabarti2006graph}. Originally, it was developed to find frequent patterns in a graph database that contains many small graphs. In this setting, the frequency (or support) is the number of graphs in the database containing the pattern. Among the most important algorithms are gSpan \cite{yan2002gspan,yan2003closegraph}, Gaston \cite{nijssen2004quickstart}, \texttt{MoSS} \cite{borgelt2007canonical}.

In single graph mining, support is redefined as the number of times a pattern appears in a single graph. However, naively counting the occurrences of a pattern breaks the anti-monotonicity requirement of the support \cite{kuramochi2005finding}: a larger pattern should have a support equal to or lower than the patterns it contains. If it does not, the search space cannot be efficiently pruned. For this reason, different definitions of support have been proposed \cite{kuramochi2005finding,fiedler2007support,bringmann2008frequent,elseidy2014grami,abdelhamid2016scalemine}.

Our paper extends the link prediction literature by extending our previous work \cite{coscia2020multiplex}. Our improved framework has better time scalability, can handle edge directions properly and, as a result, enables an improved link prediction performance. Our approach is based on \texttt{MoSS}' \cite{borgelt2005moss} ability to perform pattern mining on multiplex networks, networks where nodes can be connected by multiple qualitatively different links \cite{berlingerio2011foundations,kivela2014multilayer,boccaletti2014structure,dickison2016multilayer}. MoSS was originally developed for simple graphs -- i.e. with nodes connected by at most one edge and a single attribute per edge --, but during development we discovered that its canonical forms methodology can naturally extend to multiplex graphs. This enables it for our use and potentially more -- in the case one has multiple attributes per edge. Multiplex networks have been widely adopted in a variety of network analysis applications such as community discovery \cite{mucha2010community,berlingerio2011finding}, node ranking \cite{de2015ranking}, spreading processes \cite{de2016physics}, and even probabilistic motif analysis \cite{battiston2017multilayer}.

To the best of our knowledge, there are only three approaches that come close to multiplex graph pattern mining, each with its own downside: (i) a special case with only two layers \cite{bachi2012classifying} (signed networks), (ii) FANMOD \cite{wernicke2006fanmod}, included in Muxviz \cite{de2015muxviz}, which uses a non-monotonic support definition, and (iii) a subgraph mining approach \cite{anchuri2018mining}, which requires to provide the input patterns of interest -- and also has a non-monotonic support definition. None of these limitations apply to our proposed approach.

\subsection{Link Prediction}
In link prediction we observe a network at different moments in its evolution. The task is to estimate the likelihood of appearance of unobserved links \cite{liben2007link,lu2011link,zhang2018link}. Most link predictors determine the link likelihood either using topological properties of the network -- thus they are unable to predict links connecting existing nodes to hitherto unobserved nodes --, and/or operate on networks where nodes can connect to each other only via the same type of relation. Our approach has neither limitation.

First, we do not use topological measures, but we extract network motifs and we use them to build graph association rules. We base this part of our methodology on \texttt{GERM} \cite{berlingerio2009mining,bringmann2010learning}, and we improve over it by considering multiplex networks.

Second, we tackle multiplex link prediction, to predict the link type connecting two nodes \cite{rossetti2011scalable}. Although there are many approaches to this problem \cite{pujari2015link,jalili2017link,sharma2016efficient,hristova2016multilayer,de2017community}, they share the general idea of combining single layer scoring functions to consider inter-layer correlations.

Link prediction can be done via graph embedding techniques \cite{goyal2018graph}. We know of one multilayer graph embedding technique \cite{li2018multi} which has not been used for multiplex link prediction, and another \cite{matsuno2018mell} which has -- and we use the latter as a state-of-the-art comparison to our framework.

\subsection{Applications}
Link prediction in general, and multiplex link prediction specifically, has a number of applications in many fields. Here we briefly discuss its relevance to the network analysis community. One key aspect of online social media is the multiple identities of the same individuals across different platforms. To truly understand the spreading of information and social behaviors online, one needs to align social networks across platforms \cite{zhang2015multiple}: to identify the same users having profiles on Facebook, Twitter, etc. Some platforms might be harder to crawl than others, thus one could use multiplex link prediction to complete the information in one layer by extracting relevant rules from the other layers.

Other applications of multiplex link prediction for online social media include the analysis of brokerage between individuals in virtual and in-presence social networks \cite{hristova2015multilayer}; inform researchers on the privacy risks associated with the complex structural information embedded in multilayer networks \cite{rossi2015k}; and the planning of cross-platform marketing campaigns \cite{vikatos2020marketing}.

\section{Problem Definition}

\subsection{Directed Multiplex Network Model}

\begin{figure}
\centering
\includegraphics[width=.6\columnwidth]{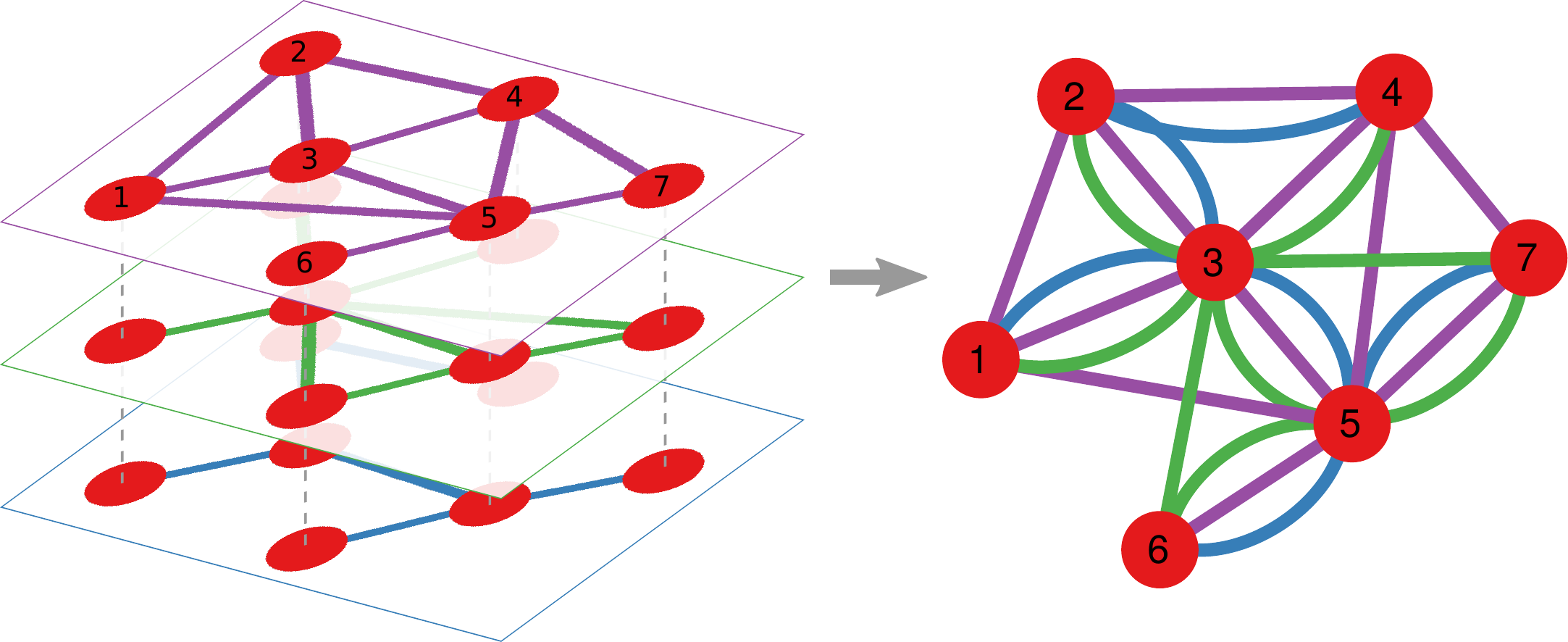}
\caption{Transforming a multilayer network with one-to-one couplings (left) into its corresponding labeled multigraph (right).}
\label{fig:model}
\end{figure}

Our model is a multiplex network, which is a labeled multigraph. Multiplex networks are equivalent to multilayer networks with one-to-one inter-layer couplings, meaning that there is a one-to-one correspondence between nodes in different layers -- as Figure \ref{fig:model} shows.

Formally, a directed multiplex network is a quadruple $G = (V, L, E, A)$, where: $V$ is the set of nodes; $L$ is the set of link labels; $E$ is the set of multiplex links, i.e.~triplets $(u,v,l)$, with $u,v \in V$ and $l \in L$. The network is directed, thus $(u,v,l) \neq (v,u,l)$; $A$ is the set of categorical node attribute values -- which we use following previous works showing their usefulness in describing social status \cite{leskovec2010predicting}. Each node has a single attribute value $a \in A$.

\subsection{Multiplex Link Prediction}
Let us assume that $G_t$ represents the status of the multiplex graph $G$ at time $t$. Given two times, $t'$ and $t''$, with $t' < t''$, we expect $G_{t'} \neq G_{t''}$. Specifically, we assume that a certain set of links were added to $E_{t'}$. Our model could be extended in a straightforward way to cover the possibility of disappearing links \cite{noel2011unfriending} but for simplicity we follow traditional link prediction and only focus on the links that were added to $E_{t'}$. Specifically, we have a target set of links defined as $T = E_{t''}-E_{t'}$, the set of all links in $E_{t''}$ but not in $E_{t'}$. 

\begin{figure}
\centering
\includegraphics[width=.6\columnwidth]{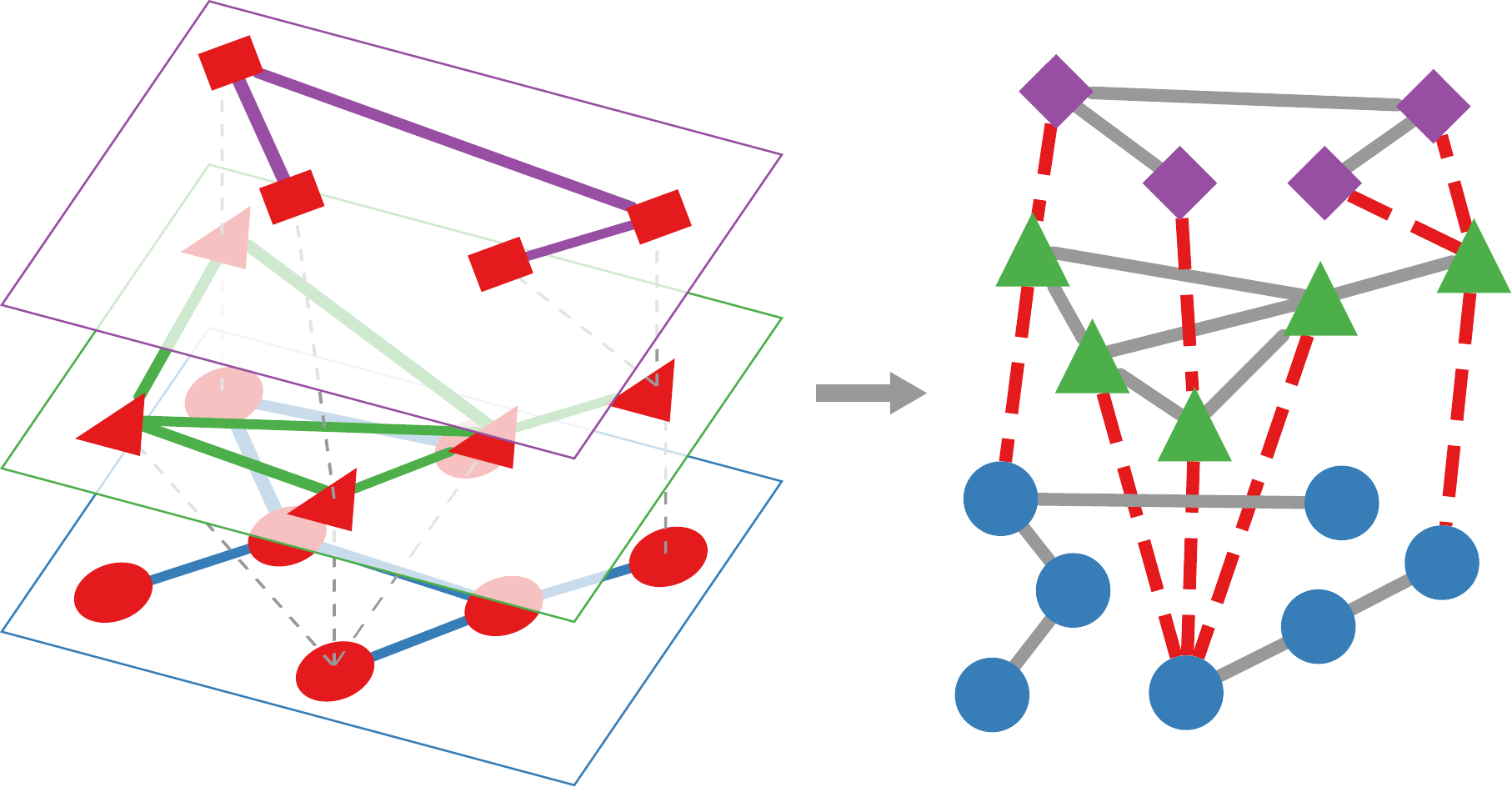}
\caption{Transforming a multilayer network with many-to-many couplings (left) into its corresponding labeled multigraph (right).}
\label{fig:many-to-many}
\end{figure}

\begin{figure*}
\centering
\includegraphics[width=.6\textwidth]{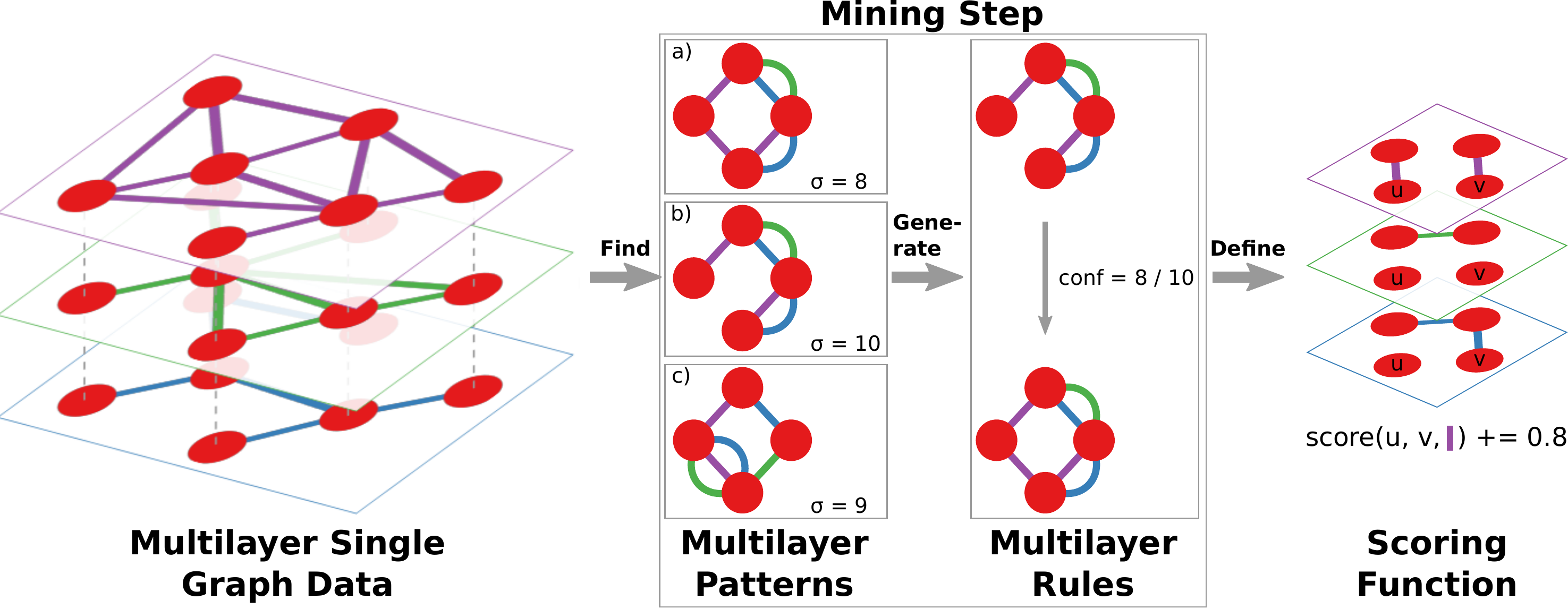}
\caption{The multilayer graph association rule mining framework.}
\label{fig:magma}
\end{figure*}

The link prediction problem is to estimate a $score(u,v)$ function for every missing link $(u,v) \not\in E_{t'}$. The $score(u,v)$ function should rate highly the missing links that are most likely to be part of $T$. In multiplex link prediction, the $score$ function takes an additional parameter: the link type~$l$. Thus, our aim is to estimate $score(u,v,l)$, for every $(u,v,l) \not\in E_{t'}$. Since the multilayer network is directed, $score(u,v,l) \neq score(v,u,l)$.

\subsection{Extension to Many-to-Many Multilayer Networks}
While this paper focuses on multiplex networks, it is possible to use our framework to perform many-to-many multilayer network mining. This is achieved by adding a pre- and post-processor to transform the data.

Figure \ref{fig:many-to-many} illustrates the procedure. In the pre-processing phase, the multilayer graph is transformed into a simple graph with two edge types: links of type $1$ are inter-layer coupling, while nodes of type $2$ are regular intra-layer connections. Each node is labeled with the layer in which it appears. The post-processing phase undoes the pre-processing. Any frequent pattern found on the simple graph contains all the information to reconstruct the original multilayer pattern. Links of type $1$ connect the different identities of the same node across layers. Links of type $2$ are intra-layer connections and one can reconstruct to which layer they belong by looking at the layer information from the node label.

We choose to ignore this extension for the rest of the paper because it affects the interpretation of one of the parameters of our framework.

\section{Methods}

\begin{figure}
\centering
\begin{subfigure}{.3\columnwidth}
\includegraphics[width=\textwidth]{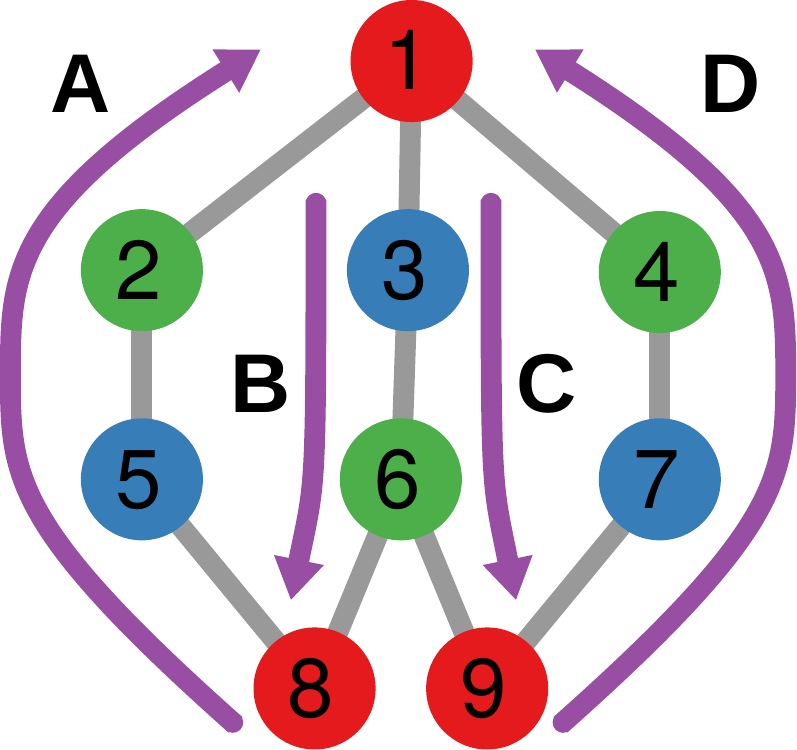}
\caption{}
\end{subfigure}\qquad
\begin{subfigure}{.04\columnwidth}
\includegraphics[width=\textwidth]{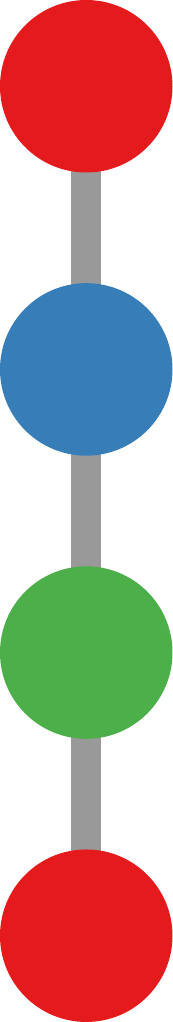}
\caption{}
\end{subfigure}\qquad
\begin{subfigure}{.4\columnwidth}
  \begin{tabular}{cccc|c}
    $A$ & $B$ & $C$ & $D$ & Count \\
    \hline
    $8$ & $1$ & $1$ & $9$ & $3$\\
    $5$ & $3$ & $3$ & $7$ & $3$\\
    $2$ & $6$ & $6$ & $4$ & $3$\\
    $1$ & $8$ & $9$ & $1$ & $3$\\
  \end{tabular}
\caption{}
\end{subfigure}
\caption{(a) The original graph, nodes labeled by their id and colored according to their label. Arrows indicate and label the occurrences of the pattern in the graph. (b) The pattern. (c) The image table for the minimum image support definition, with the pattern's nodes as rows and all possible occurrences of the pattern as columns. Each cell records the node id we use for the mapping.}
\label{fig:mis}
\end{figure}

\subsection{The Framework}
Figure~\ref{fig:magma} shows an overview of our framework. First, we find frequent multiplex graph patterns using \texttt{MoSS}. \texttt{MoSS} enables the use of multiple support definitions, of which we use the minimum number of node images, since it is the closest to the actual number of occurrences of a pattern \cite{bringmann2008frequent}. As mentioned above, \texttt{MoSS} allows to find frequent patterns in a single graph, and it accepts labeled multi-graphs as inputs. The support definition determines the frequency of a pattern by counting the number of different nodes in the original graph that can play a specific role in the pattern, and taking the minimum. Figure~\ref{fig:mis} provides an example: there are four ways to map Figure~\ref{fig:mis}(b) in Figure~\ref{fig:mis}(a), but its support is three because we have to re-use the same node in the same role for some of these occurrences.

The second step is building the set $R$ of multiplex graph association rules. In contrast to the old framework, this step happens inside \texttt{MoSS}' pattern mining phase. This avoids the old framework's step of re-matching all mined patterns' embeddings on the input structure and it is the basis of most runtime gains.

\texttt{MoSS} requires three parameters: minimum support $\sigma$, maximum pattern size $s$, and minimum confidence $c$. The minimum support is the minimum number of occurrences of the pattern to be considered frequent and included in the results. The maximum pattern size is the maximum number of nodes in a pattern. The minimum confidence is the value of confidence below which a rule is discarded.

To keep complexity and potential overfitting under control, we decide to focus exclusively on rules which predict the appearance of a single new link. In other words, we build a $p_1 \rightarrow p_2$ rule if pattern $p_2$ completely includes $p_1$, with a single additional edge -- and, potentially, an additional node (see Section \ref{sec:method-oldnew}). Moreover, we exclusively focus on connected rules, i.e. neither the antecedent nor the consequent have more than one connected component. Both $p_1$ and $p_2$ need to be frequent patterns, appearing more than $\sigma$ times in $G$. The single link differentiating $p_2$ from $p_1$ tells us the two nodes we expect to be connected and the link type. The weight of the rule is its confidence: the ratio of the support of the consequent over the support of the antecedent. If this weight is below $c$, the rule is discarded. Due to the anti-monotonicity of support, the confidence of a rule cannot be greater than 1: $p_2$ is larger than $p_1$, hence its support can only be equal or less.

Every time we encounter $p_1$ in $G$, we can identify the two nodes and the type of the missing link by looking at all its $p_2$ consequents in $R$.

In practice, $score(u, v, l)$ is the count of all rules saying $u$ should connect to $v$ in $l$, weighted by their confidence. There could be multiple weighting schemes -- simple count, lift, average confidence, etc. In previous work, we decided to focus on a confidence-weighted score, showing that count, confidence sum, and lift sum all perform equivalently and are interchangeable, while average confidence and average lift perform erratically and should be ignored. For this paper, we follow previous work and we weight rules by their confidence sum.

\subsection{Predicting Old-New Links}\label{sec:method-oldnew}
New links can attach to nodes that were not part of the network: $V_{t'} \subseteq V_{t''}$. There are two node classes in $V_{t''}$: ``old'' nodes which are nodes in $V_{t'}$, and ``new'' nodes from $V_{t''} - V_{t'}$. Each link in $E_{t''} - E_{t'}$ can belong in one of three categories: ``old-old'' links connect two old nodes, ``old-new'' links connect an old node with a new one, and ``new-new'' links connect two new nodes.

\begin{figure}
\centering
\includegraphics[width=.6\columnwidth]{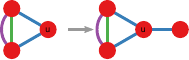}
\caption{A rule allowing us to predict old-new links. The link color represents its type.}
\label{fig:oldnew2}
\end{figure}

Traditional link prediction exclusively deals with scoring old-old links where all scoring functions need to be calculated on the topology of $G_{t'}$, thus nodes not in $G_{t'}$ cannot contribute to $score(u,v)$. On the contrary, here we are able to predict old-new links by exploiting the rules in which the consequent has one node more than the antecedent. To see how this is possible consider the rule in Figure \ref{fig:oldnew2}. Consequents are matched to antecedents if they contain them, minus one link. The new link is free to connect to an additional node, not necessarily to a node that was already part of the antecedent. Using the rule in the figure, we can predict that node $u$ will connect to a previously unobserved node.

With the restriction we impose -- that the rules can only contain a single additional edge -- it is not possible to predict new-new links. It would be in principle possible if we were to allow two or more new edges in the consequent, but we leave this investigation for future work.

\subsection{Adapting \texttt{MoSS}}
As pointed out in Section \ref{sec:related-fpm}, we discovered that \texttt{MoSS} already naturally handles multiplex graphs -- graphs with multiple edges with different labels between nodes. For this reason, we did not need to develop a new canonical form to efficiently explore the search space during the mining step.

\begin{figure}
\centering
\includegraphics[width=.6\columnwidth]{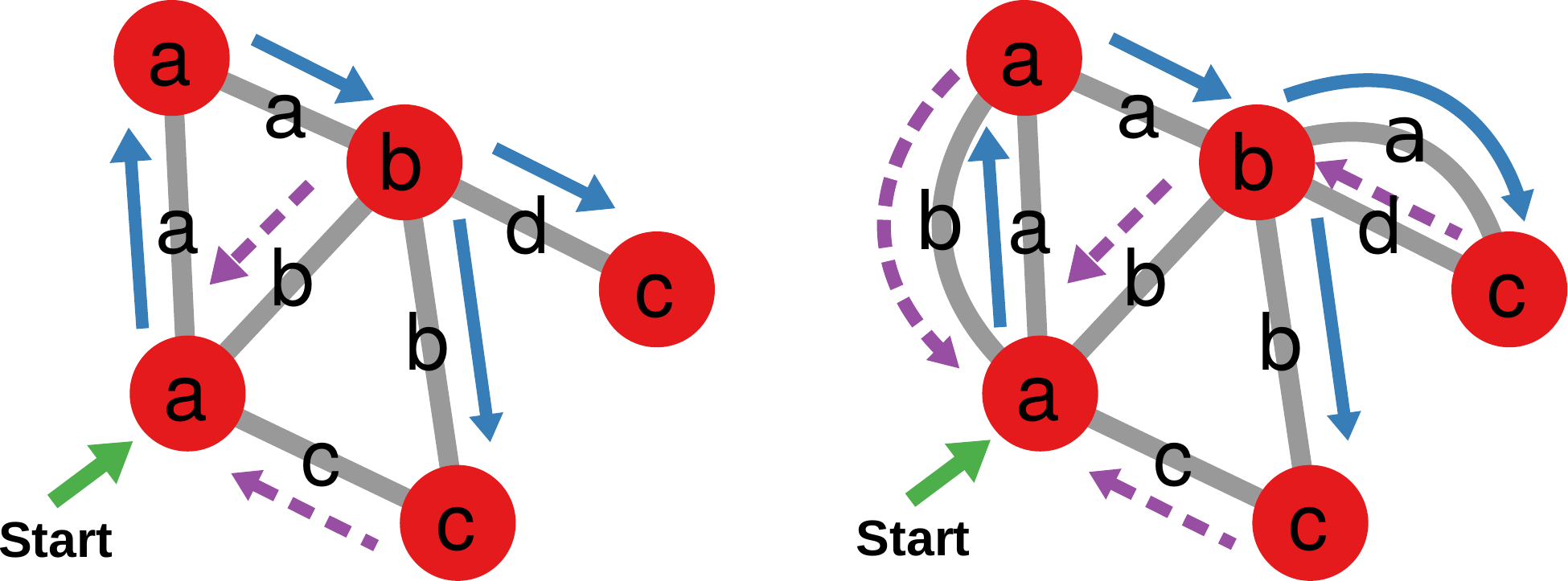}
\caption{DFS approach to build canonical codes for a simple graph (left) and multigraph (right). Green arrow points at the first explored node, solid blue arrows show the DFS tree, dashed purple arrows highlight the edges closing a cycle.}
\label{fig:canonical}
\end{figure}

The procedure  boils down to exploring a pattern via a spanning tree to construct a canonical form of a pattern, a useful construct to establish isomorphism without actually running a computationally intensive test. The spanning tree can be built using a DFS -- like in gSpan \cite{yan2002gspan} -- or a BFS -- like in \texttt{MoSS} proper \cite{borgelt2005moss} -- exploration strategy. Edges not in the spanning tree needs to be inserted into this canonical form in a consistent way. All these edges close cycles in one way or another, otherwise they would be part of the spanning tree. \texttt{MoSS} can handle multigraphs because parallel edges in a multigraph are no more special than other non-parallel cycle-closing edges, that are already naturally handled by this canonical form construction. In fact, parallel edges are simply closing a cycle of length two. Figure \ref{fig:canonical} shows an example of labeled multigraph (i.e. a multiplex graph) which requires no special handling.

In this paper we adopt the BFS strategy as default, because for the data sets we use in our study this empirically performs better -- see Section \ref{sec:exp-runtime-canonical}.

However, originally \texttt{MoSS} did not handle edge directions. In our old framework, edge directions were handled in post-process -- which induces mistakes as un-directed patterns mined by \texttt{MoSS} are interpreted as directed. Here, we have incorporated the edge direction into \texttt{MoSS} itself.

We need only one extra bit of information to encode edge direction -- since the fact that two edges might exist between the same two nodes is already handled by the ability of MoSS to work on multigraphs. Specifically, each node is assigned an id once it is discovered. Edges are then stored in a format where the lower id always precedes the higher id. We can then add a single bit of information recording whether the actual edge direction correspond to this lower-to-higher id or not.

\section{Experiments}

\subsection{Setup}

\subsubsection{Data Sets}

\begin{table}[]
\centering
\begin{tabular}{l|rrrrr}
Network & $|V|$ & $|E|$ & $|L|$ & Dir & Dyn\\
\hline
Aarhus & 61 & 620 & 5 & N & N\\
Physicians & 241 & 1,551 & 3 & Y & N\\
CElegans & 279 & 5,863 & 3 & Y & N\\
Pardus & 6,373 & 78,661 & 3 & Y & Y\\
\end{tabular}
\caption{Basic statistics of the data sets: $|V|$, number of nodes; $|E|$, number of links; $|L|$, number of layers; Dir, whether the network is directed; Dyn, whether the network has temporal information.}
\label{tab:data}
\end{table}

\textbf{Aarhus} \cite{magnani2013combinatorial} records interactions in the CS department of Aarhus University. Employees can establish five different types of relations: coauthorship, lunch, collaboration, etc. This is a static undirected network.

\textbf{Physicians} \cite{coleman1957diffusion} tracks relations between physicians asking three questions. Each physicians reports with whom they: ask advice, discuss cases, and/or have a friendship relation. Each question generates a link type in the network. This is a directed network.

\textbf{CElegans} \cite{chen2006wiring} is the neurological structure of the C.\ Elegans worm. There are three types of connections, each corresponding to a different link type: electric, chemical monadic, and chemical polyadic. This is a directed network.

\textbf{Pardus} \cite{szell2010multirelational,szell2010msd} includes relations between players from an online game.\footnote{\url{https://www.pardus.at/}} Players can be each other's friends or enemies, and can attack each other. This generates three layers, one positive (friendship) while the others (enemies and attacks) are negative. This is a temporal directed network. We use the network on day 300 as training set, and the network observed 100 days later as the test set.

We also generate synthetic data from powercluster models \cite{holme2002growing}, a synthetic graph type that can reproduce realistic features of real world networks such as a power law degree distribution and high clustering coefficient.

Table \ref{tab:data} reports basic statistics of our data sets. The reported sizes (number of nodes $|V|$, and edges $|E|$) of the data sets are the unions of their training and test sets (both the number of nodes and links might increase from training to training+test, as new nodes might be introduced). All data sets except Pardus come from the CoMuNe project\footnote{\url{https://comunelab.fbk.eu/data.php}} \cite{de2013mathematical}. We remove all loops.

\subsubsection{Baseline algorithms}
Here we briefly present the state of the art of multiplex link prediction.

Sharma \cite{sharma2016efficient} calculates the likelihood of having a link of type $l_1$ given that the nodes are connected by link type $l_2$: $p_{l_2,l_1}$. Then, 

$$score(u,v,l_1) = \sum \limits_{l_2 \in L} p_{l_2,l_1} \delta_{u,v,l_2},$$
with $\delta_{u,v,l_2}$ being equal to $1$ if nodes $u$ and $v$ are connected in $l_2$, $0$ otherwise. The downside is that every node pair not connected in any layer will get a score of zero. While this makes it the most memory efficient approach by dramatically reducing output size, it also makes it miss all connections between previously completely disconnected nodes, which routinely happen in real world networks.

Pujari \cite{pujari2015link} takes a collection of classical link prediction scores (Common Neighbor, Adamic-Adar, etc.) for each link type separately as input features for a decision tree. It adds multiplex features such as the score average and entropy across layers. A disadvantage is a lack of features for pairwise link type interactions, only for the overall interaction between all link type pairs. A related method \cite{hajibagheri2016holistic} adds temporal information, but reduces to the Pujari method for static networks, thus for our purposes they are equivalent.

Jalili \cite{jalili2017link} builds a metagraph by performing community discovery on each link type separately using Infomap \cite{rosvall2008maps}, then it counts the number of simple metapaths of length $1$, $2$, and $3$ that lead from node $u$ to node $v$ either starting or ending in layer $l$. It generates six features as the input of an SVM with a Gaussian kernel. Paths cannot contain cycles -- however, it is possible to calculate them by multiplying the adjacency matrix with itself and removing the diagonal, since the paths are capped to be of length 3.

Hristova \cite{hristova2016multilayer} calculates a series of classical scores per link type. It then generates multiplex features by aggregating these scores, and feeds them to a Random Forest classifier. The paper defines a number of features that are inapplicable here because they are tailored for special geotemporal data (Twitter and Foursquare). They also define two multiplex aggregations, which they call ``global'' and ``core''. Here we use the global one, as the core aggregation is too restrictive and leads to a too sparse output.

De Bacco \cite{de2017community} defines a multilayer mixed-membership stochastic blockmodel \cite{airoldi2008mixed} by assuming that nodes belong to the same groups across layers -- a more relaxed version has also been recently proposed \cite{roxana2019edge}. The group-group affinity is different in each layer, allowing for pairs of layers to be correlated, anti-correlated, or independent. It then finds the best node-node and group-group connection probabilities via the expectation maximization algorithm which serve as the scores for the link prediction task.

Mell \cite{matsuno2018mell} is a technique to create node embeddings -- like the ones created using DeepWalk \cite{perozzi2014deepwalk} or node2vec \cite{grover2016node2vec} -- for multilayer networks. Once the multilayer node embeddings are created, they can be used to create node-node-layer similarity scores that estimate the likelihood of the nodes connecting in a specific layer.

\subsubsection{Validation}
Some data sets have temporal information and some do not. For the data sets without temporal information, we perform the link prediction task using ten-fold cross validation as the split between training and test. We build each test set by randomly drawing 10\% of the edges, which means that it might contain nodes that are not in the training fold -- if we picked all of their edges. For the data set with temporal information, we collect data until time $t$ for the training data, and we use data starting from time $t$ until $t + \delta$ for the test.

\subsubsection{Machine}
We run all experiments on a Intel Xeon W-11955M, running Ubuntu 22.04, with 32GB of RAM. The only exception is the Pardus data, for which we allocate 192GB of RAM instead.

\subsection{Runtime Gains}

\subsubsection{Changing Parameters}
\begin{figure*}
\centering
\begin{subfigure}{.45\columnwidth}
\includegraphics[width=\columnwidth]{{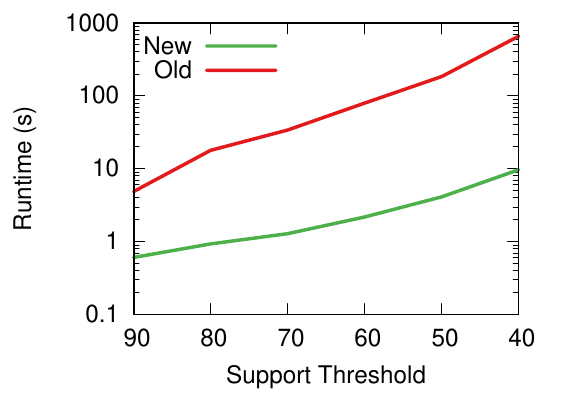}}
\caption{Support}
\end{subfigure}\qquad
\begin{subfigure}{.45\columnwidth}
\includegraphics[width=\columnwidth]{{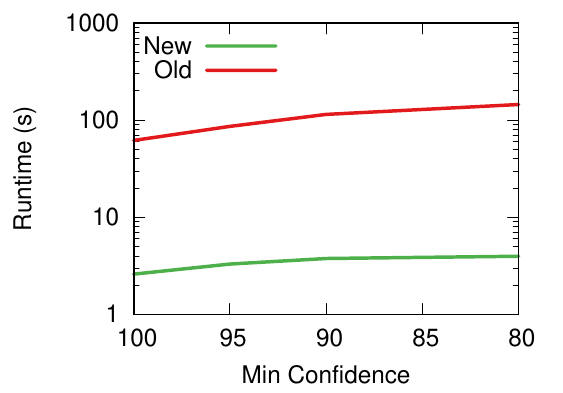}}
\caption{Confidence}
\end{subfigure}\qquad
\begin{subfigure}{.45\columnwidth}
\includegraphics[width=\columnwidth]{{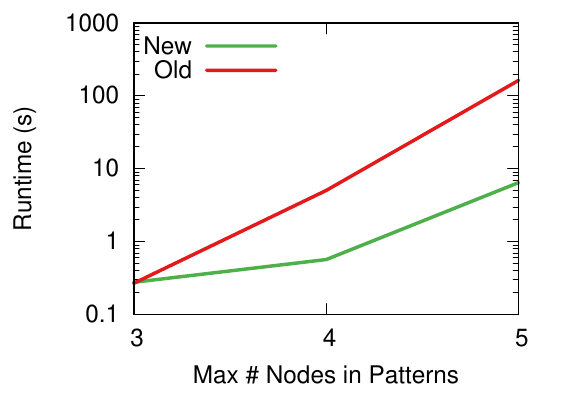}}
\caption{Size}
\end{subfigure}
\caption{The running time (y axis) as we modify one parameter (x axis) for the old framework (red) and the new one (green).}
\label{fig:runtimes-1}
\end{figure*}

In this section we compare the new multiplex association rule mining framework with the old one, in terms of processing time. We start by testing the effect of different choices in parameters, specifically support, confidence, and maximum pattern size. This test was performed on the Aarhus network. Figure \ref{fig:runtimes-1} reports the results.

In Figure \ref{fig:runtimes-1}(a) we change the minimum support. The higher the minimum support, the shorter the algorithm takes, as we are pruning the search space more aggressively. This is what we see, as the new version of the framework takes half a second to run with support at 90\%, while around 10 seconds with support at 40\%. The improvement over the old framework is noticeable: not only it is faster by one order of magnitude at low support levels, but its asymptotic behavior is better. At 40\% support, the new framework is two orders of magnitude faster than the old one.

In Figure \ref{fig:runtimes-1}(b) we change the minimum confidence. The higher the confidence, the shorter the algorithm takes, because we have fewer rules to apply as we discard the ones with low confidence. This is what we see, as the new version of the framework takes 2.4 seconds to run with confidence at 100\%, while around 3.6 seconds with confidence at 80\% -- a 50\% runtime increase. Not only the new framework faster than the old one by one order of magnitude at low support levels, but its asymptotic behavior is better. At 100\% confidence, the new framework is 20 times faster than the old one, while at 80\% confidence it is 40 times faster.

In Figure \ref{fig:runtimes-1}(c) we change the maximum pattern size in number of nodes. Larger pattern size requires an exponentially larger search space, affecting runtime. This is what we see, as the new version of the framework takes a quarter of a second to find patterns of size 3, while around 6 seconds for patterns of size 5. The improvement over the old framework is noticeable: for patterns of size 3 the runtimes are indistinguishable, because there are so few patterns that the algorithm terminates effectively instantaneously. For patterns of size 5, the new code is almost 30 times faster.

\begin{figure}
\centering
\includegraphics[width=.5\columnwidth]{{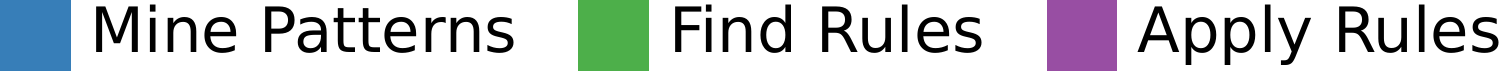}}\\
\begin{subfigure}{.45\columnwidth}
\includegraphics[width=\columnwidth]{{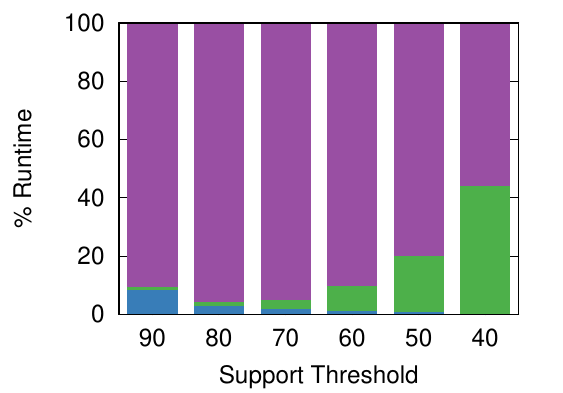}}
\caption{Old}
\end{subfigure}\quad
\begin{subfigure}{.45\columnwidth}
\includegraphics[width=\columnwidth]{{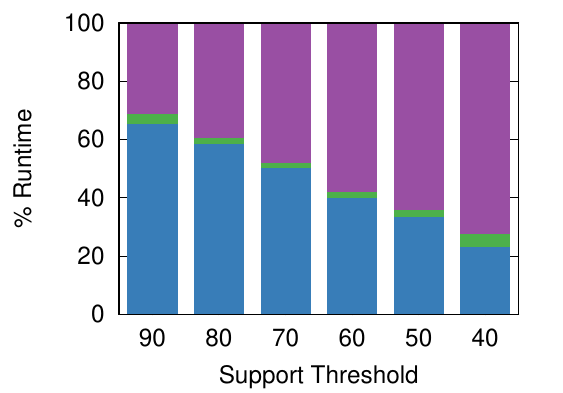}}
\caption{New}
\end{subfigure}
\caption{The breakdown of runtime (y axis) for different support thresholds (x axis).}
\label{fig:runtimes-2}
\end{figure}

Figure \ref{fig:runtimes-2} shows where the improvement lies. In the figure, we break down the running time into the three main phases of the framework -- we ignore the preprocessing as its contribution to runtime is irrelevant. In the old framework (Figure \ref{fig:runtimes-2}(a)), most of the time is taken by the application of the rules at high support values, and by finding and applying the rules for low support values. This is because these two steps are done as postprocess, after the pattern mining phase has finished.

On the other hand, the new framework (Figure \ref{fig:runtimes-2}(b)) can find rules while mining the patterns, and it can do so more efficiently -- because every pattern extension is automatically recognized as a potential rule. A significant portion of the runtime is now taken by the (fast) mining step, and the application of the rules is also more efficient -- thus it does not impact runtime as much even if it constitutes most of the computation for low support values.

One important thing to note is that we report here single core performance. However, in the rule applying step, each rule is applied independently, and thus this step can be made massively parallel. This is yet another advantage for the new framework: as the support threshold goes down, the new framework spends more and more time on a step that can be rendered much more efficient by parallelizing it, while the old framework's runtime instead gets dominated by a step that cannot -- the one in which we derive the rules from the mined frequent patterns.

\subsubsection{Changing Input Size}
It is also interesting to know how runtime is affected by different topologies. For this we need to abandon the Aarhus network and generate a set of synthetic graphs. In these graphs we can control the number of nodes, the number of layers, the average degree, and the number of node labels. 

\begin{figure}
\centering
\begin{subfigure}{.45\columnwidth}
\includegraphics[width=\columnwidth]{{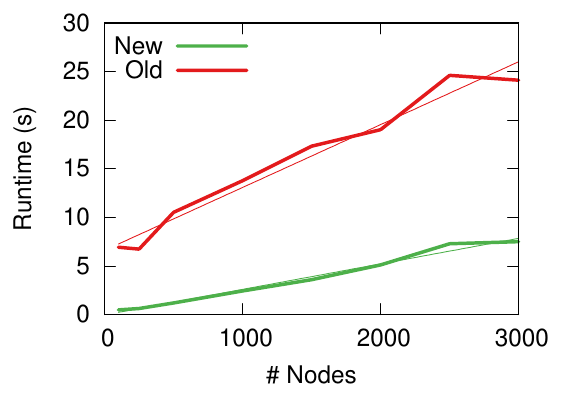}}
\caption{$|V|$}
\end{subfigure}\quad
\begin{subfigure}{.45\columnwidth}
\includegraphics[width=\columnwidth]{{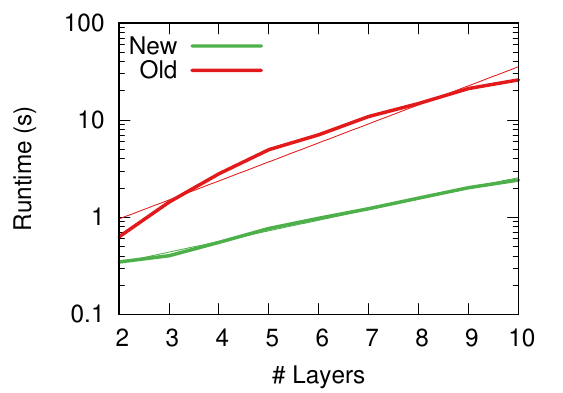}}
\caption{$|L|$}
\end{subfigure}
\begin{subfigure}{.45\columnwidth}
\includegraphics[width=\columnwidth]{{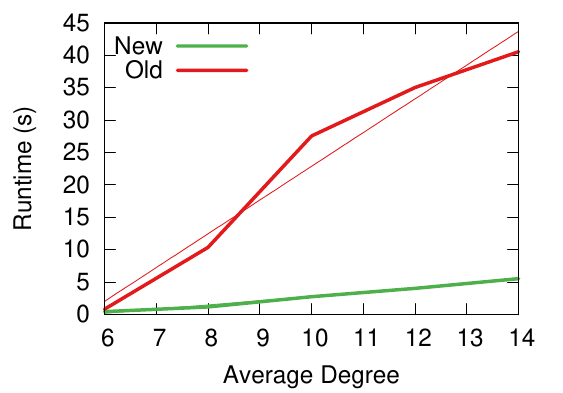}}
\caption{$\bar{k}$}
\end{subfigure}\quad
\begin{subfigure}{.45\columnwidth}
\includegraphics[width=\columnwidth]{{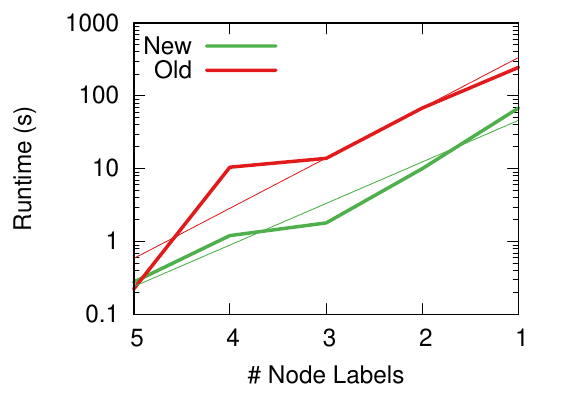}}
\caption{Node Labels}
\end{subfigure}
\caption{The runtime (y axis) of the new (green) and old (red) framework when changing different topological values (x axis) in synthetic networks.}
\label{fig:runtimes-3}
\end{figure}

Figure \ref{fig:runtimes-3} shows the results, including a trendline, as the running time might vary depending on the randomization of the network. For each experiment, we repeat the procedure ten times and take the average running time. The default parameters for the tests are: number of nodes at 500, average degree at~8, the number of layers at~7, and the number of node labels at~4. For each test we modify one of these parameters leaving the others constant.

In the first experiment (Figure~\ref{fig:runtimes-3}(a)), we increase the number of nodes from 100 to 3000. We can see that the new framework has both an offset and a trend advantage. The slope of the (linear) trendline is 0.002 for the new framework against 0.006 for the old one, meaning that the new framework scales better by a factor of~3.

Moving to Figure~\ref{fig:runtimes-3}(b), we change the number of layers from 2 to 10. In this case, the trendline is exponential, with exponent 0.26 for the new framework and 0.46 for the old one, showing a substantial runtime gain.

In Figure~\ref{fig:runtimes-3}(c), we increase the average degree from~6 to~14. Again, the new framework has a trend advantage. The slope of the (linear) trendline is 0.64 for the new framework against 5.22 for the old one, meaning that the new framework scales better by a factor of~8.

Finally, Figure~\ref{fig:runtimes-3}(d) focuses on changing the number of node labels from~5 to~1. Also in this case we have an exponential fit. In this case, the asymptotic behavior of the new framework is the same as the old one, as the exponent of the fit does not change. The multiplicative speedup represented by the constant is still relevant, as the new framework is an entire order of magnitude faster.

Overall, the new framework proved to be significantly faster than the old one in all tests performed.

\subsubsection{BFS vs DFS Pattern Expansion}\label{sec:exp-runtime-canonical}

\begin{figure}
\centering
\begin{subfigure}{.45\columnwidth}
\includegraphics[width=\columnwidth]{{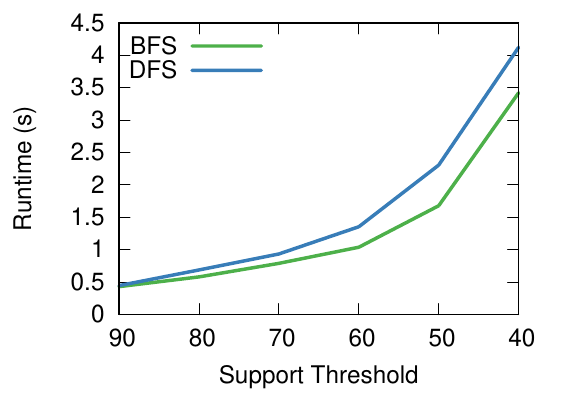}}
\caption{Aarhus}
\end{subfigure}\quad
\begin{subfigure}{.45\columnwidth}
\includegraphics[width=\columnwidth]{{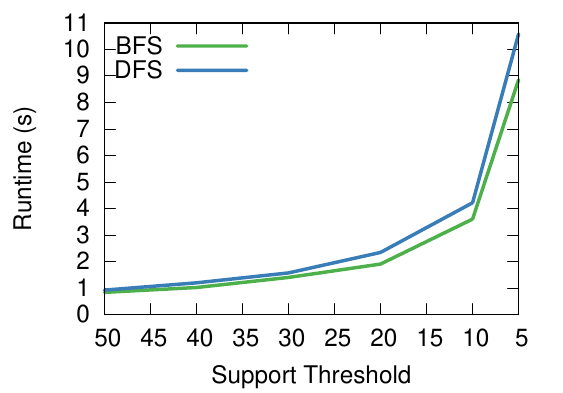}}
\caption{Phys}
\end{subfigure}
\caption{The runtime (y axis) of the BFS (green) and DFS (blue) pattern expansion techniques for growing support (x axis).}
\label{fig:runtimes-4}
\end{figure}

In Figure \ref{fig:runtimes-4} we compare the running times of the algorithm when using a BFS and a DFS approach to expanding the patterns. For space reason we only show the results on the Aarhus and the Phys data sets, noting that they are representative of all other tested networks. Specifically, BFS has a slight advantage over DFS, although it is usually pretty marginal. However, that justifies our choice of using BFS as the default for this paper.

\subsection{Performance}
We test the performance of multiplex graph association rules against the state of the art on all networks. We use the standard approach of building a ROC curve and calculating the area under the curve (AUC) as evaluation. Since we test a lot of methods, to avoid overcrowding the plots we only show four methods. We always show the old and new versions of the framework (in red and green respectively). We also always show Mell (in blue) as it is the only deep learning algorithm we use. Finally, we include (in purple) the Best Traditional Competitor (BTC): this is the method with the highest AUC among all the non-graph mining non-deep learning alternatives.

\subsubsection{Static Multiplex Networks}
\begin{figure}
\centering
\begin{subfigure}{.45\columnwidth}
\includegraphics[width=\columnwidth]{{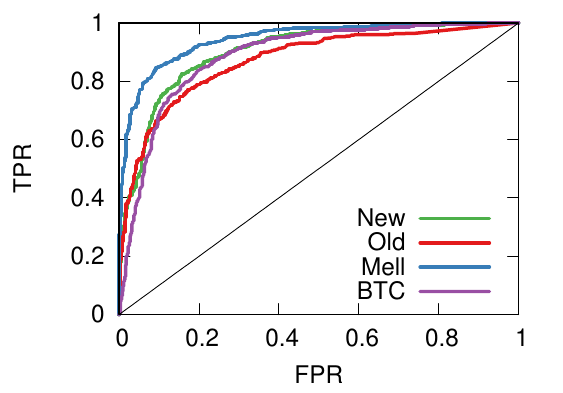}}
\caption{Aarhus}
\end{subfigure}\quad
\begin{subfigure}{.45\columnwidth}
\includegraphics[width=\columnwidth]{{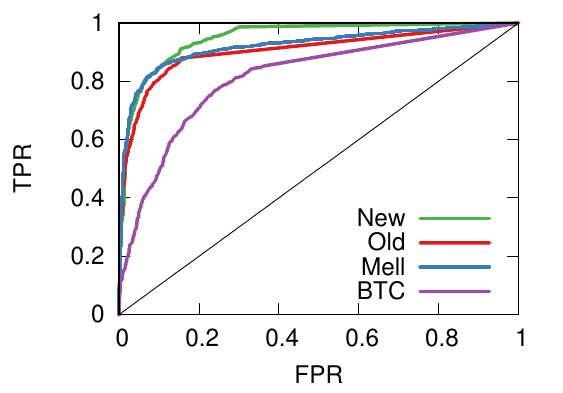}}
\caption{Physicians}
\end{subfigure}
\begin{subfigure}{.45\columnwidth}
\includegraphics[width=\columnwidth]{{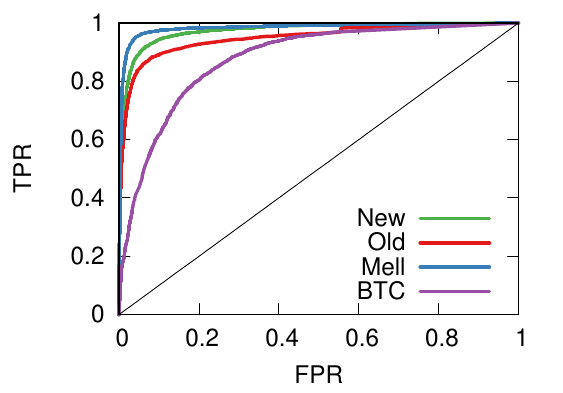}}
\caption{C. Elegans}
\end{subfigure}\quad
\begin{subfigure}{.45\columnwidth}
\includegraphics[width=\columnwidth]{{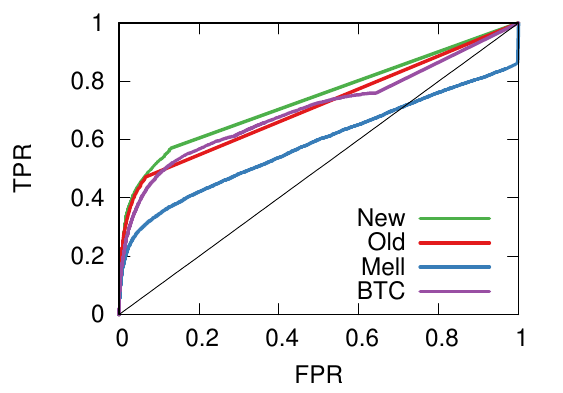}}
\caption{Pardus}
\end{subfigure}
\caption{The ROC curves for all tested data sets. BTC = Best Traditional Competitor.}
\label{fig:roc-ml}
\end{figure}

Figure~\ref{fig:roc-ml} shows the ROC curves for all the methods on all data sets.  Table \ref{tab:aucs} reports the corresponding AUC values.

\begin{table}
\footnotesize
\centering
\begin{tabular}{l|rrrr}
Method & Aarhus & Physicians & CElegans & Pardus\\
\hline
\textbf{New} & 0.903 & \textbf{0.949} & 0.974 & \textbf{0.741}\\
\textit{Old} & 0.871 & 0.905 & 0.949 & 0.711\\
Mell         & \textbf{0.945} & 0.922 & \textbf{0.983} & 0.584\\
Sharma       & 0.804 & 0.735 & 0.677 & 0.511\\
Pujari       & 0.854 & 0.637 & 0.810 & 0.611\\
Jalili       & 0.887 & 0.800 & 0.685 & 0.640\\
Hristova     & 0.885 & 0.745 & 0.755 & 0.509\\
De Bacco     & 0.881 & 0.814 & 0.878 & 0.704\\
\hline
Ens. Base    & 0.940 & 0.945 & 0.980 & 0.765\\
Ens. Opt.    & 0.944 & 0.946 & 0.984 & 0.800\\
\end{tabular}
\caption{AUC of the ROC curves from Figure \ref{fig:roc-ml}. Highest performance values among the non-ensemble methods in bold.}
\label{tab:aucs}
\end{table}

The Aarhus data set is small with few edges that can be predicted. For this reason, most methods perform excellently at roughly the same level in terms of AUC. A notable exception is Mell, which performs significantly better than everything else. The high performance of the traditional methods can be explained by the fact that the Aarhus network is small and it is simple enough to be approached with traditional methods. The ROC curve (Figure \ref{fig:roc-ml}(a)) confirms what we see from the table.

In the Physicians data set we see one of the clearest advantages of the new framework over the old one in terms of performance. The increase in efficiency allows us to specify a lower support threshold. This, in turn, allow us to retrieve more patterns with the same runtime cost. More patterns means more rules, which enable the prediction of a set of edges that were left unscored in the old framework -- when the ROC curve straightens in Figure \ref{fig:roc-ml}(b) it means that the predictor is starting to make random guesses for the edges that it would all equally score to zero. The old framework was beaten by Mell, but our updated version is significantly better than any alternative, Mell included.

The C.Elegans data set is where both graph mining and Mell perform the best, eclipsing all traditional methods. The high performance of Mell can be explained because the data set is large enough for a proper neural network training phase, but it still does not contain any old-new type of links to be predicted, which are invisible to Mell. Figure \ref{fig:roc-ml}(c) confirms that there is no significant difference between Mell and our new framework, both well above the best traditional competitor.

\subsubsection{Dynamic Multiplex Network}
The Pardus data set, being the largest, most complex, and the only one with actual temporal information on the edges, is the most important and interesting test case. Among the traditional predictors, De Bacco performs extremely well. It performs much better than Mell. Both methods are blind to old-new type links. However, Mell assigns a score to all links in the test set, while De Bacco will leave some of those scores at zero. For this reason, Mell is penalized to worse-than-chance level for the hardest links to predict.

The old version of the framework is on par with De Bacco's performance. However, the added efficiency of the new version of the framework allows us to mine more rules. This implies an increased precision, which makes the new version of the framework the winner in terms of AUC.

\begin{figure}
\centering
\begin{subfigure}{.45\columnwidth}
\includegraphics[width=\columnwidth]{{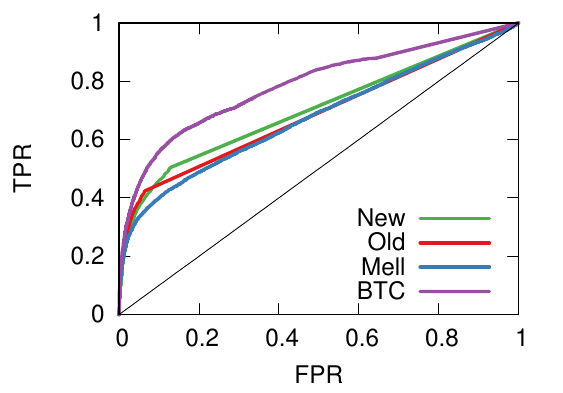}}
\caption{Old-old}
\end{subfigure}\quad
\begin{subfigure}{.45\columnwidth}
\includegraphics[width=\columnwidth]{{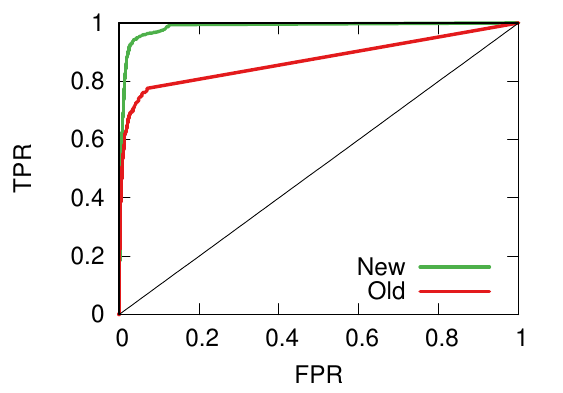}}
\caption{Old-new}
\end{subfigure}
\caption{The ROC curves for Pardus on the old-old and old-new segments of the test set. BTC = Best Traditional Competitor.}
\label{fig:roc-ml-oldold}
\end{figure}

\begin{table}
\footnotesize
\centering
\begin{tabular}{l|rrr}
Method & Pardus Old-Old & Pardus Old-New\\
\hline
\textbf{New} & 0.703 & \textbf{0.986}\\
\textit{Old} & 0.686 & 0.872\\
Mell         & 0.676 & ?.???\\
Sharma       & 0.513 & ?.???\\
Pujari       & 0.667 & ?.???\\
Jalili       & 0.740 & ?.???\\
Hristova     & 0.511 & ?.???\\
De Bacco     & \textbf{0.787} & ?.???\\
\hline
Ens. Base    & 0.798 & ?.???\\
Ens. Opt.    & 0.808 & ?.???\\
\end{tabular}
\caption{AUC of the ROC curves from Figure \ref{fig:roc-ml-oldold}. Highest performance values among the non-ensemble methods in bold. ?.???: method could not provide a score.}
\label{tab:aucs-oldold}
\end{table}

Predicting old-new links is challenging, but it is crucial to understand the evolution of a growing network. This alone is a reason to use a pattern mining approach over the alternatives. Yet, it is informative to understand how the methods perform on both old-new and old-old links. To do so, we investigate what happens to the performance of each method once we remove all old-new links from the test set. We do so and recalculate the ROC curves in Figure \ref{fig:roc-ml-oldold}(a) and their AUCs in Table \ref{tab:aucs-oldold} (first column).

All baselines increase their AUC, which is expected, because we remove links they cannot predict, for which their performance is random (or worse-than-random for Mell). Our framework, instead, lowers its AUC, because it spreads its predictive power on the harder problem of predicting old-old and old-new links at the same time. If one is exclusively interested in predicting old-old links, the best choice would be the De Bacco predictor.

If we instead focus only on old-new links -- ROC curves in Figure \ref{fig:roc-ml-oldold}(b) and their AUCs in Table~\ref{tab:aucs-oldold} (second column) --, we can see how the new framework relates to the old. The new framework outperforms the old both on old-old links, but especially on old-new links. In the latter case, the increase in performance is remarkable, showing how powerful the new framework is in predicting the growth of a network in terms of new nodes incoming. If this is the focus of one's research, not only is there no alternative to using our framework, but its performance is also nigh optimal.

\subsubsection{Ensemble Classifier}
We combine all classifiers tested into an ensemble classifier incorporating all scores. Ensemble classifiers combine scores from all classifiers, which usually increases overall performance. It is useful to assess how much a method could be improved by adding more information.

Our ensemble classifier works in two steps. First, it normalizes the scores of the methods so that their average equals zero and their standard deviation equals one. This way, all classifier scores are on the same scale. Second, it searches via basin hopping for the best weighting score, i.e.\ the one maximizing prediction quality, by multiplying each predictor score by a weight. This ensemble classifier is not overfitted, because the basin hopping phase is performed by splitting the training set into an internal training and test. However, note that, if the network is small, splitting an already small training set into parts might lower the ensemble's performance, which might result in lower scores. We use the ``Ensemble Opt.'' label in Table~\ref{tab:aucs} to refer to the optimized ensemble. Without the simulated annealing step, equally weighting all methods, the ensemble has a lower AUC -- ``Ensemble Base'' in Table \ref{tab:aucs}.

``Ensemble Opt.'' has comparable performance with the best performing algorithm on all networks. The scenario in which an ensemble classifier would be most useful is on Pardus, since De Bacco performs well on old-old links and our framework performs well for old-new links. Combining the two could lead to the best of both worlds -- and, in fact, that is the network for which we observe the biggest gains from the ensemble.

\subsubsection{Monoplex Networks}
To show the effectiveness of our approach also on traditional single layer networks, we create single layer views of our data set by collapsing all edges across all layers in a single one. In this case, two nodes are connected if they are connected in any of the layers. The exception is Pardus for which we only keep the friendship edges -- since it would not make sense to consider friendships and enmities as the same type of link.

Since this is merely a feasibility study, we compare with classical single layer link prediction scores such as Resource Allocation \cite{zhou2009predicting} (RA), Jaccard \cite{liben2007link} (JA, a variant of common neighbors), Preferential Attachment \cite{liben2007link} (PA), and Adamic-Adar \cite{adamic2003friends} (AA). We also use the node2vec link prediction implementation \cite{grover2016node2vec}, to compare our approach with graph neural networks using node embeddings -- node2vec being a single layer equivalent of Mell.

\begin{figure}
\centering
\begin{subfigure}{.45\columnwidth}
\includegraphics[width=\columnwidth]{{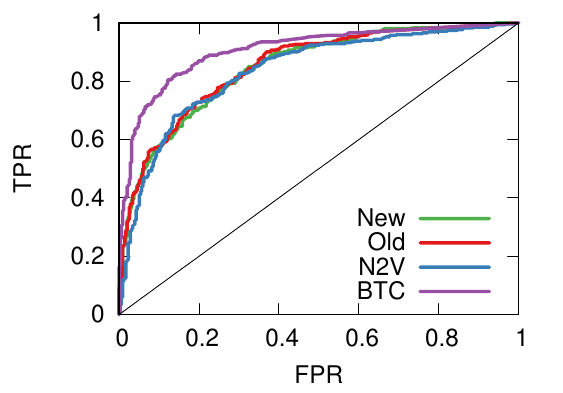}}
\caption{Aarhus}
\end{subfigure}\quad
\begin{subfigure}{.45\columnwidth}
\includegraphics[width=\columnwidth]{{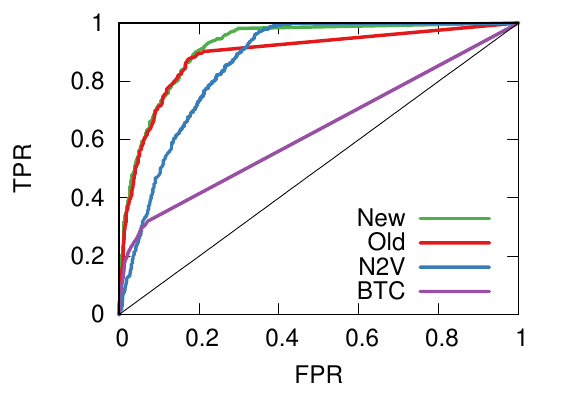}}
\caption{Physicians}
\end{subfigure}
\begin{subfigure}{.45\columnwidth}
\includegraphics[width=\columnwidth]{{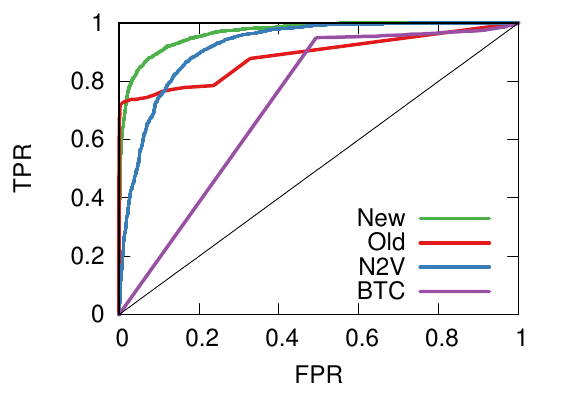}}
\caption{C. Elegans}
\end{subfigure}\quad
\begin{subfigure}{.45\columnwidth}
\includegraphics[width=\columnwidth]{{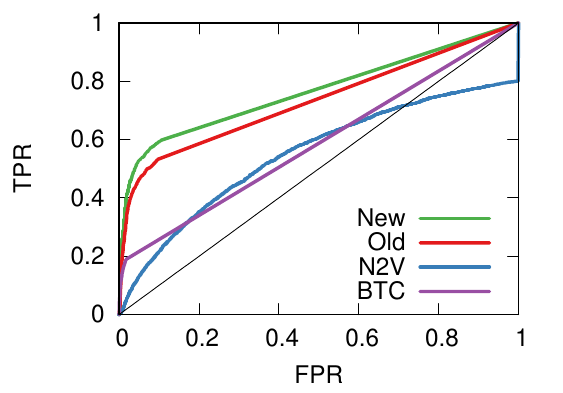}}
\caption{Pardus}
\end{subfigure}
\caption{The ROC curves for all tested data sets on single layer networks. BTC = Best Traditional Competitor.}
\label{fig:roc-sl}
\end{figure}

\begin{table}
\footnotesize
\centering
\begin{tabular}{l|rrrr}
Method & Aarhus & Physicians & CElegans & Pardus\\
\hline
\textbf{New} & 0.847 & \textbf{0.921} & \textbf{0.967} & \textbf{0.765}\\
\textit{Old} & 0.852 & 0.891 & 0.887 & 0.732\\
RA           & \textbf{0.907} & 0.629 & 0.600 & 0.586\\
PA           & 0.662 & 0.540 & 0.723 & 0.507\\
JA           & 0.903 & 0.629 & 0.601 & 0.586\\
AA           & 0.901 & 0.629 & 0.600 & 0.586\\
n2v          & 0.836 & 0.865 & 0.921 & 0.549\\
\hline
Ens. Base    & 0.901 & 0.909 & 0.947 & 0.696\\
Ens. Opt.    & 0.912 & 0.923 & 0.974 & 0.760\\
\end{tabular}
\caption{AUC of the ROC curves from Figure \ref{fig:roc-sl}. Highest performance values in bold.}
\label{tab:aucs-single}
\end{table}

Figure \ref{fig:roc-sl} shows the resulting ROC curves and Table~\ref{tab:aucs-single} their AUCs. The results can be interpreted as follows.

First, RA, JA, and AA are variants of common neighbors: they focus on the same limited set of potential links, with slightly different scores. They are largely interchangeable.

Second, given that RA, JA, and AA only consider potential links between nodes at two hops away, they work especially well in extremely small and dense networks like Aarhus. Their performance degrades quickly as the networks grows in size. If the network is as small and dense as Aarhus, there is no need to use a complex method such as the one we present in this paper.

Third, with the exception of Aarhus, we still see a large difference between the improved version of the algorithm and the one presented in \cite{coscia2020multiplex}. This can be explained by our improved way of dealing with directed networks. In fact, Aarhus is the only network that is not directed, and there is no difference in performance there.

Finally, our approach is superior to performing link prediction using the node2vec embeddings.

\section{Case Study}
For our case study, we follow our previous work \cite{coscia2020multiplex} and investigate the Pardus network more deeply. The objective of this section is twofold: first we want to show how the new improved algorithm is able to uncover patterns that were previously unavailable. Second, using these patterns, we can give a more complete picture of some of the salient characteristics of the Pardus network.

\subsection{New Patterns}
In our previous case study \cite{coscia2020multiplex}, we showed how the pattern mining approach could extend social balance theories by considering patterns of four nodes, rather than being limited to triangles. The increased efficiency of the framework now allows to find even more complex patterns. Here we report two rules of five nodes among those we could find, using comparable computational resources as the ones used in the previous study.

\begin{figure}
\centering
\begin{subfigure}{.4\columnwidth}
\includegraphics[width=\columnwidth]{{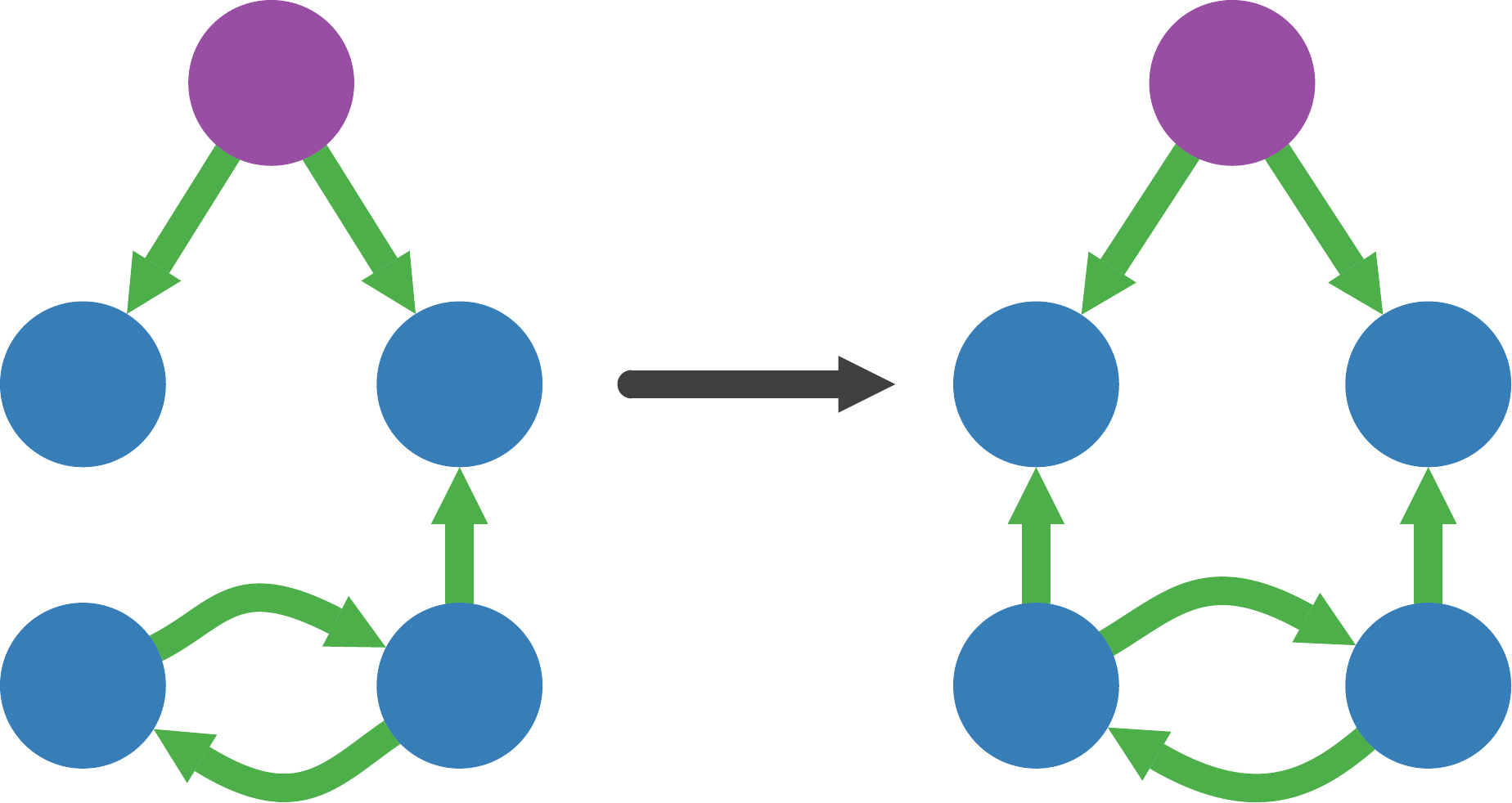}}
\caption{Rule \#544}
\end{subfigure}\qquad
\begin{subfigure}{.4\columnwidth}
\includegraphics[width=\columnwidth]{{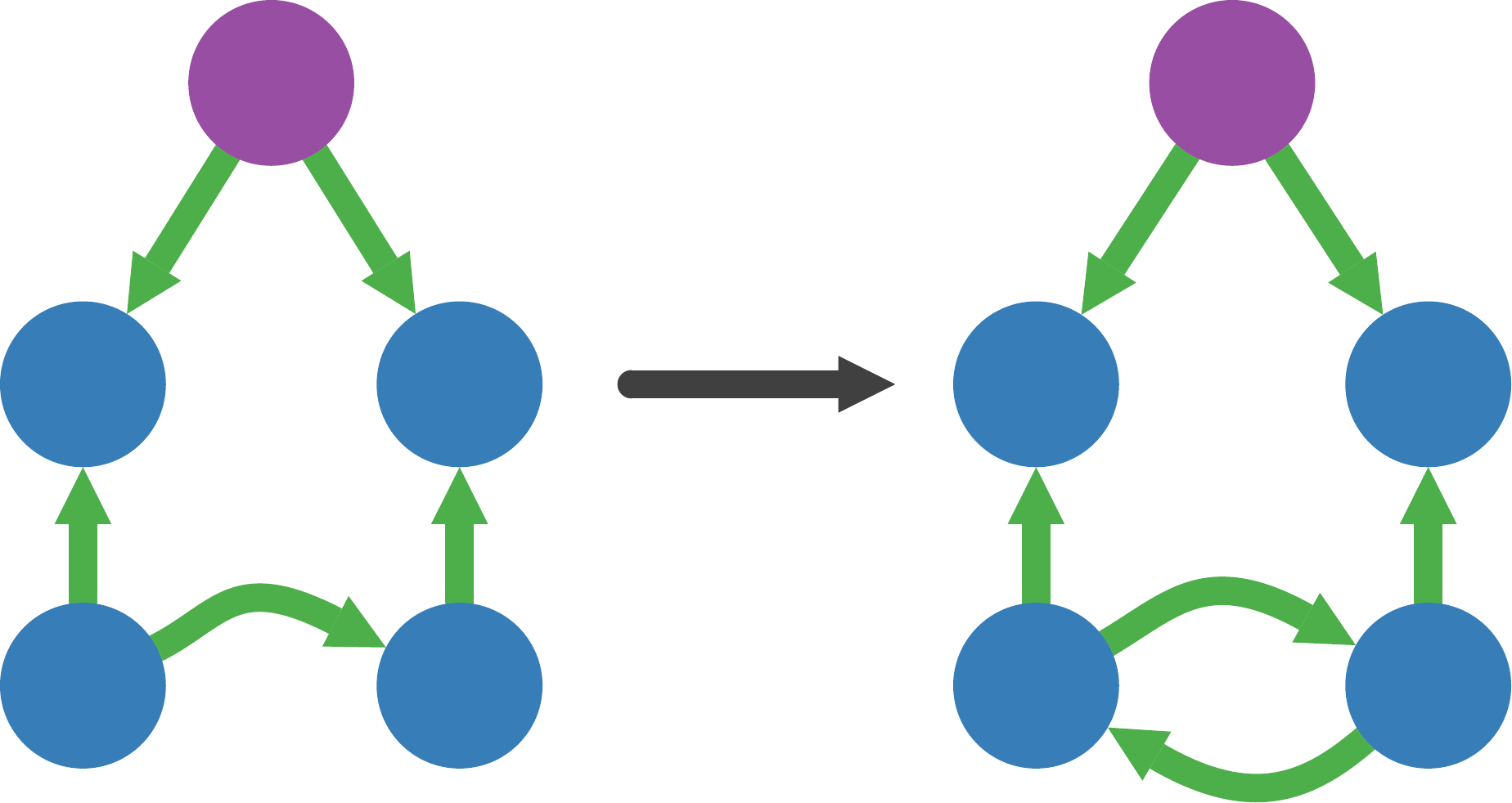}}
\caption{Rule \#573}
\end{subfigure}
\caption{Two frequent rules with five nodes in the Pardus network (minimum support = 19\%). Nodes in purple are power users (top quartile of game experience), nodes in blue are regular ones. Green edges show friendships.}
\label{fig:pardus-5}
\end{figure}

Figure \ref{fig:pardus-5} shows these two rules. In both cases, we see the social network expanding via the addition of friendship connections. In Figure \ref{fig:pardus-5}(a) we see a large pattern being closed, while in Figure \ref{fig:pardus-5}(b) we see a connection in the pattern being reciprocated -- note that the consequent is the same in both rules. These patterns of five nodes were hitherto unfeasible to be mined before the developments described in this paper.

All the rules of five nodes we have found include only friendship connections. There are no single attack or enmity edges in patterns of five nodes, showing how friendship is the most common cause for relationships among a large set of users. ``Wars'' involving attacks and enmities between large groups of users exist, but they are much rarer than large groups of friends and thus would require extremely low support thresholds to be found.

\subsection{Frustration Analysis}
With the ability of going beyond triangles, we can provide a more detailed analysis of social balance: the tendency of signed social networks to follow specific connection patterns -- e.g. ``friend of my friend is my friend'' \cite{heider1958psychology, antal2005dynamics, leskovec2010predicting, szell2010multirelational, kirkley2019balance}. For each pattern, we can calculate its frustration \cite{harary1959measurement, aref2019balance}.

To estimate frustration, a two-step procedure is required \cite{aref2016exact}. First, patterns are partitioned into two groups, in such a way that within a group there are only positive connections, and between groups there are only negative connections. This group detection might not always be possible: some edges are frustrated -- they show either negative relationships within a group or positive relationships between groups. The frustration index is the minimum share of edges one would need to remove (or sign flip) to successfully detect coherent groups.

Since we can estimate the frustration of each pattern, we can also classify rules on whether they increase, decrease, or maintain frustration. Frustration-increasing rules are interesting: they break the expectation of social balance theory.

\begin{figure}
\centering
\begin{subfigure}{.4\columnwidth}
\includegraphics[width=\columnwidth]{{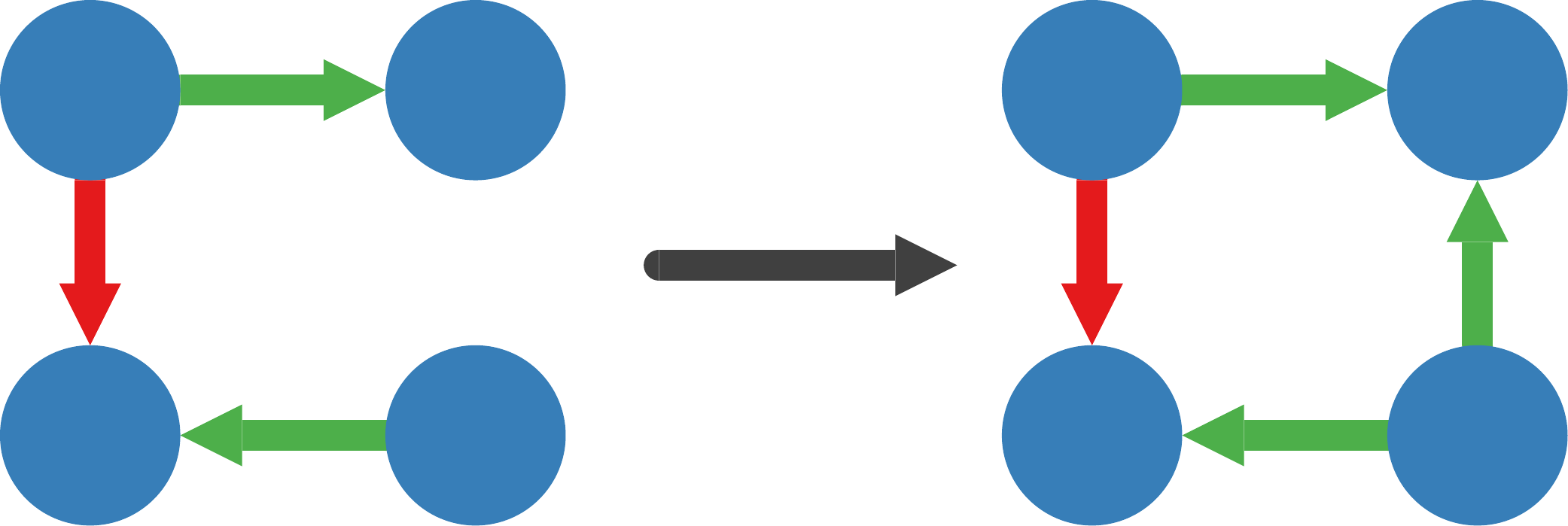}}
\caption{Rule \#3423}
\end{subfigure}\qquad
\begin{subfigure}{.4\columnwidth}
\includegraphics[width=\columnwidth]{{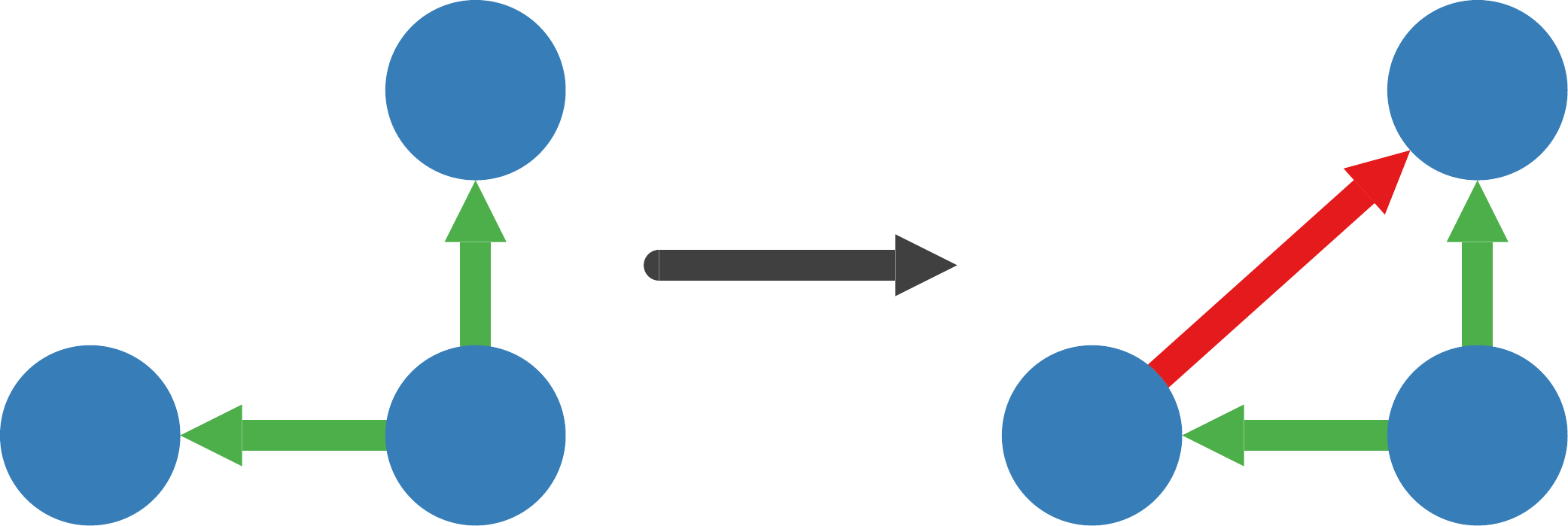}}
\caption{Rule \#279}
\end{subfigure}
\caption{Two frustration-increasing rules in the Pardus network (minimum support = 6\%). Nodes in blue are regular users. Green edges show friendships, red edges show enmities.}
\label{fig:pardus-frustration-rules}
\end{figure}

Figure \ref{fig:pardus-frustration-rules} shows two interesting frustration-increasing rules. Figure \ref{fig:pardus-frustration-rules}(a) has the highest confidence among these rules: 84\% of the times, a user will befriend another user that is friend with a friend's enemy -- which is unexpected according to social balance theory. Figure \ref{fig:pardus-frustration-rules}(b) is one of the rules that induces the most frustration in the network, which is the creation of an unbalanced triangle. Noticeably, this rule has very low confidence, only 31\% of triads with two positive edges will close with a negative one.

While we could find some frustration-increasing rules with the previous framework, we are now able to perform a more in-depth statistical analysis by sampling a larger number of them. Specifically, we can use this analysis to confirm that the Pardus society largely conforms to the expectation of social balance. We base this conclusion on the following observations.

\begin{figure}
\centering
\begin{subfigure}{.45\columnwidth}
\includegraphics[width=\columnwidth]{{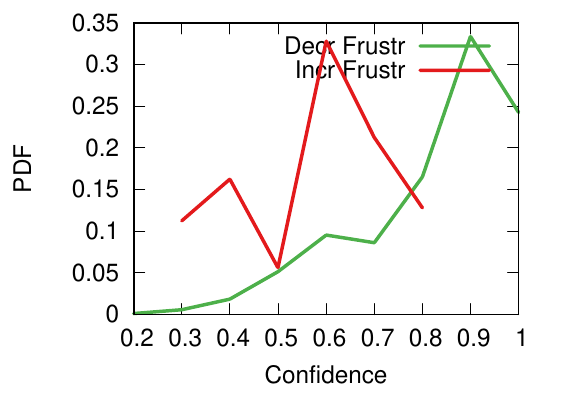}}
\caption{Confidence}
\end{subfigure}\quad
\begin{subfigure}{.45\columnwidth}
\includegraphics[width=\columnwidth]{{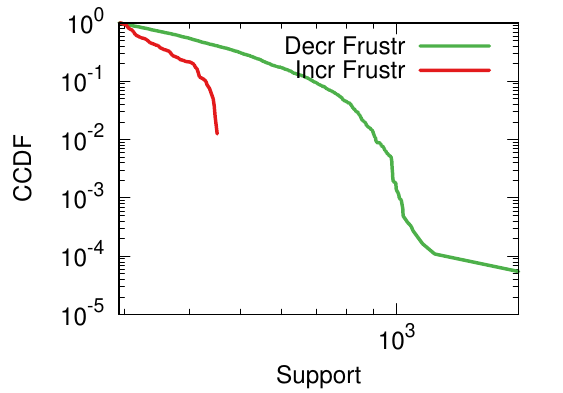}}
\caption{Support}
\end{subfigure}
\caption{The PDF of confidence (a), and CCDF of support (b) of frustration inducing (red) and frustration decreasing (green) rules in the Pardus network.}
\label{fig:pardus-frustration-distr}
\end{figure}

Only 1.7\% of rules increase frustration. On the other hand, 97.5\% of rules result in a pattern with zero frustration.

The remaining 0.8\% rules have an antecedent with non-zero frustration and they all result in a consequent with strictly lower frustration than the antecedent. This decrease of frustration prompts the interesting question for future research on the role of forgiveness in social systems \cite{laifa2015overview}.

The average confidence of frustration-increasing rules is lower than the average confidence of rules resulting in zero frustration (57.4\% vs 82.5\%). Figure~\ref{fig:pardus-frustration-distr}(a) confirms that, while there is no frustration-increasing rule with confidence higher than $\sim80\%$, most rules with zero frustration consequents have a confidence of $\sim90\%$.

The frustration-increasing rules also have lower average support (347 vs 461), meaning they are not only with lower confidence and in lower count, but they are also less frequently found in the data. Figure~\ref{fig:pardus-frustration-distr}(b) confirms that, while there is no frustration-increasing rule with support higher than $\sim450$, $40\%$ of rules with zero frustration have a support higher than that, with a maximum support of $1,712$, almost four times as high.

Note that in Figure~\ref{fig:pardus-frustration-distr}(a) we report the probability density function of the confidence of a rule, i.e. the share of rules with a given confidence value. Figure \ref{fig:pardus-frustration-distr}(b) reports the complementary cumulative distribution function instead: the share of rules with a given value of support or higher.

This deeper analysis was made possible by the computational efficiency of the new framework and goes into deeper details than limiting the analysis to triangles.

\section{Conclusion}
In this paper, we describe a new framework to perform multiplex link prediction via graph association rules. Multiplex link prediction is the task of forecasting new links appearing in a multiplex network, specifying not only which two nodes will connect to each other, but also with which type. We perform a series of experiments showing how this approach can match the state of the art using graph neural networks -- and vastly outperforms it in the larger, more complex scenario that also requires predicting incoming nodes in the network. This latter aspect is a qualitative improvement we bring to multiplex link prediction, which was previously understudied. 

This work is an improvement over our previous work. Specifically, we improve the time efficiency and we have a better way of handling directed connections. These improvements result in quantitative performance gains in link prediction, but also qualitative ones. In the signed network scenario, we extend classical social balance theory by considering patterns of five nodes, rather than limiting to patterns of four nodes as in our previous work.

There are a number of future directions to further increase multiplex link prediction performance. We could perform the experiments on the extended framework with many-to-many interlayer mappings, which we outlined. Second, we could investigate more scoring schemes rather than relying on the simple rule count weighted by confidence. Despite these possibilities for technical improvements, we proved the usage of graph association rules to be a quantitative and qualitative improvement over previous multiplex link predictors, with unique domain applications. With this new extension, we also made this approach usable in real world scenarios.

\section*{Acknowledgements}
Christian Borgelt gratefully acknowledges the financial support from Land Salzburg within the WISS 2025 project IDA-Lab (20102-F1901166-KZP and 20204-WISS/225/197-2019).

\bibliographystyle{IEEEtran}
\bibliography{IEEEabrv,biblio.bib}

\begin{thebibliography}{10}
\providecommand{\url}[1]{#1}
\csname url@samestyle\endcsname
\providecommand{\newblock}{\relax}
\providecommand{\bibinfo}[2]{#2}
\providecommand{\BIBentrySTDinterwordspacing}{\spaceskip=0pt\relax}
\providecommand{\BIBentryALTinterwordstretchfactor}{4}
\providecommand{\BIBentryALTinterwordspacing}{\spaceskip=\fontdimen2\font plus
\BIBentryALTinterwordstretchfactor\fontdimen3\font minus
  \fontdimen4\font\relax}
\providecommand{\BIBforeignlanguage}[2]{{%
\expandafter\ifx\csname l@#1\endcsname\relax
\typeout{** WARNING: IEEEtran.bst: No hyphenation pattern has been}%
\typeout{** loaded for the language `#1'. Using the pattern for}%
\typeout{** the default language instead.}%
\else
\language=\csname l@#1\endcsname
\fi
#2}}
\providecommand{\BIBdecl}{\relax}
\BIBdecl

\bibitem{liben2007link}
D.~Liben-Nowell and J.~Kleinberg, ``The link-prediction problem for social
  networks,'' \emph{Journal of the American society for information science and
  technology}, vol.~58, no.~7, pp. 1019--1031, 2007.

\bibitem{lu2011link}
L.~L{\"u} and T.~Zhou, ``Link prediction in complex networks: A survey,''
  \emph{Physica A}, vol. 390, no.~6, pp. 1150--1170, 2011.

\bibitem{zhang2018link}
M.~Zhang and Y.~Chen, ``Link prediction based on graph neural networks,''
  \emph{NeurIPS}, vol.~31, 2018.

\bibitem{berlingerio2009mining}
M.~Berlingerio, F.~Bonchi, B.~Bringmann, and A.~Gionis, ``Mining graph
  evolution rules,'' in \emph{ECML PKDD}.\hskip 1em plus 0.5em minus
  0.4em\relax Springer, 2009, pp. 115--130.

\bibitem{bringmann2010learning}
B.~Bringmann, M.~Berlingerio, F.~Bonchi, and A.~Gionis, ``Learning and
  predicting the evolution of social networks,'' \emph{IEEE Intelligent
  Systems}, vol.~25, no.~4, pp. 26--35, 2010.

\bibitem{coscia2020multiplex}
M.~Coscia and M.~Szell, ``Multiplex graph association rules for link
  prediction,'' \emph{ICWSM}, 2021.

\bibitem{krackhardt1987cognitive}
D.~Krackhardt, ``Cognitive social structures,'' \emph{Social networks}, vol.~9,
  no.~2, pp. 109--134, 1987.

\bibitem{roethlisberger1939management}
F.~Roethlisberger and W.~Dickson, ``Management and the worker.'' 1939.

\bibitem{kivela2014multilayer}
M.~Kivel{\"a}, A.~Arenas, M.~Barthelemy, J.~P. Gleeson, Y.~Moreno, and M.~A.
  Porter, ``Multilayer networks,'' \emph{Journal of complex networks}, vol.~2,
  no.~3, pp. 203--271, 2014.

\bibitem{boccaletti2014structure}
S.~Boccaletti, G.~Bianconi, R.~Criado, C.~I. Del~Genio, J.~G{\'o}mez-Gardenes,
  M.~Romance, I.~Sendina-Nadal, Z.~Wang, and M.~Zanin, ``The structure and
  dynamics of multilayer networks,'' \emph{Physics Reports}, vol. 544, no.~1,
  pp. 1--122, 2014.

\bibitem{berlingerio2011foundations}
M.~Berlingerio, M.~Coscia, F.~Giannotti, A.~Monreale, and D.~Pedreschi,
  ``Foundations of multidimensional network analysis,'' in \emph{ASONAM}.\hskip
  1em plus 0.5em minus 0.4em\relax IEEE, 2011, pp. 485--489.

\bibitem{dickison2016multilayer}
M.~E. Dickison, M.~Magnani, and L.~Rossi, \emph{Multilayer social
  networks}.\hskip 1em plus 0.5em minus 0.4em\relax Cambridge University Press,
  2016.

\bibitem{rossetti2011scalable}
G.~Rossetti, M.~Berlingerio, and F.~Giannotti, ``Scalable link prediction on
  multidimensional networks,'' in \emph{ICDM Workshop}.\hskip 1em plus 0.5em
  minus 0.4em\relax IEEE, 2011, pp. 979--986.

\bibitem{matsuno2018mell}
R.~Matsuno and T.~Murata, ``Mell: effective embedding method for multiplex
  networks,'' in \emph{Companion Proceedings of the The Web Conference 2018},
  2018, pp. 1261--1268.

\bibitem{pujari2015link}
M.~Pujari and R.~Kanawati, ``Link prediction in multiplex networks.''
  \emph{NHM}, vol.~10, no.~1, pp. 17--35, 2015.

\bibitem{jalili2017link}
M.~Jalili, Y.~Orouskhani, M.~Asgari, N.~Alipourfard, and M.~Perc, ``Link
  prediction in multiplex online social networks,'' \emph{Royal Society open
  science}, vol.~4, no.~2, p. 160863, 2017.

\bibitem{sharma2016efficient}
S.~Sharma and A.~Singh, ``An efficient method for link prediction in weighted
  multiplex networks,'' \emph{Computational social networks}, vol.~3, no.~1,
  p.~7, 2016.

\bibitem{hristova2016multilayer}
D.~Hristova, A.~Noulas, C.~Brown, M.~Musolesi, and C.~Mascolo, ``A multilayer
  approach to multiplexity and link prediction in online geo-social networks,''
  \emph{EPJ Data Science}, vol.~5, no.~1, p.~24, 2016.

\bibitem{de2017community}
C.~De~Bacco, E.~A. Power, D.~B. Larremore, and C.~Moore, ``Community detection,
  link prediction, and layer interdependence in multilayer networks,''
  \emph{PRE}, vol.~95, no.~4, p. 042317, 2017.

\bibitem{borgelt2005moss}
C.~Borgelt, T.~Meinl, and M.~Berthold, ``Moss: a program for molecular
  substructure mining,'' in \emph{OSDM Workshop}.\hskip 1em plus 0.5em minus
  0.4em\relax ACM, 2005, pp. 6--15.

\bibitem{chakrabarti2006graph}
D.~Chakrabarti and C.~Faloutsos, ``Graph mining: Laws, generators, and
  algorithms,'' \emph{ACM CSUR}, vol.~38, no.~1, p.~2, 2006.

\bibitem{yan2002gspan}
X.~Yan and J.~Han, ``gspan: Graph-based substructure pattern mining,'' in
  \emph{2002 IEEE International Conference on Data Mining, 2002.
  Proceedings.}\hskip 1em plus 0.5em minus 0.4em\relax IEEE, 2002, pp.
  721--724.

\bibitem{yan2003closegraph}
------, ``Closegraph: mining closed frequent graph patterns,'' in
  \emph{SIGKDD}.\hskip 1em plus 0.5em minus 0.4em\relax ACM, 2003, pp.
  286--295.

\bibitem{nijssen2004quickstart}
S.~Nijssen and J.~N. Kok, ``A quickstart in frequent structure mining can make
  a difference,'' in \emph{SIGKDD}.\hskip 1em plus 0.5em minus 0.4em\relax ACM,
  2004, pp. 647--652.

\bibitem{borgelt2007canonical}
C.~Borgelt, ``Canonical forms for frequent graph mining,'' in \emph{Advances in
  Data Analysis}.\hskip 1em plus 0.5em minus 0.4em\relax Springer, 2007, pp.
  337--349.

\bibitem{kuramochi2005finding}
M.~Kuramochi and G.~Karypis, ``Finding frequent patterns in a large sparse
  graph,'' \emph{Data mining and knowledge discovery}, vol.~11, no.~3, pp.
  243--271, 2005.

\bibitem{fiedler2007support}
M.~Fiedler and C.~Borgelt, ``Support computation for mining frequent subgraphs
  in a single graph.'' in \emph{MLG}.\hskip 1em plus 0.5em minus 0.4em\relax
  Citeseer, 2007.

\bibitem{bringmann2008frequent}
B.~Bringmann and S.~Nijssen, ``What is frequent in a single graph?'' in
  \emph{PAKDD}.\hskip 1em plus 0.5em minus 0.4em\relax Springer, 2008, pp.
  858--863.

\bibitem{elseidy2014grami}
M.~Elseidy, E.~Abdelhamid, S.~Skiadopoulos, and P.~Kalnis, ``Grami: Frequent
  subgraph and pattern mining in a single large graph,'' \emph{VLDB Endowment},
  vol.~7, no.~7, pp. 517--528, 2014.

\bibitem{abdelhamid2016scalemine}
E.~Abdelhamid, I.~Abdelaziz, P.~Kalnis, Z.~Khayyat, and F.~Jamour, ``Scalemine:
  scalable parallel frequent subgraph mining in a single large graph,'' in
  \emph{SC Conference}.\hskip 1em plus 0.5em minus 0.4em\relax IEEE Press,
  2016, p.~61.

\bibitem{mucha2010community}
P.~J. Mucha, T.~Richardson, K.~Macon, M.~A. Porter, and J.-P. Onnela,
  ``Community structure in time-dependent, multiscale, and multiplex
  networks,'' \emph{science}, vol. 328, no. 5980, pp. 876--878, 2010.

\bibitem{berlingerio2011finding}
M.~Berlingerio, M.~Coscia, and F.~Giannotti, ``Finding and characterizing
  communities in multidimensional networks,'' in \emph{ASONAM}.\hskip 1em plus
  0.5em minus 0.4em\relax IEEE, 2011, pp. 490--494.

\bibitem{de2015ranking}
M.~De~Domenico, A.~Sol{\'e}-Ribalta, E.~Omodei, S.~G{\'o}mez, and A.~Arenas,
  ``Ranking in interconnected multilayer networks reveals versatile nodes,''
  \emph{Nature communications}, vol.~6, p. 6868, 2015.

\bibitem{de2016physics}
M.~De~Domenico, C.~Granell, M.~Porter, and A.~Arenas, ``The physics of
  spreading processes in multilayer networks,'' \emph{Nature Physics}, vol.~12,
  no.~10, p. 901, 2016.

\bibitem{battiston2017multilayer}
F.~Battiston, V.~Nicosia, M.~Chavez, and V.~Latora, ``Multilayer motif analysis
  of brain networks,'' \emph{Chaos: An Interdisciplinary Journal of Nonlinear
  Science}, vol.~27, no.~4, p. 047404, 2017.

\bibitem{bachi2012classifying}
G.~Bachi, M.~Coscia, A.~Monreale, and F.~Giannotti, ``Classifying
  trust/distrust relationships in online social networks,'' in
  \emph{SocialCom}.\hskip 1em plus 0.5em minus 0.4em\relax IEEE, 2012, pp.
  552--557.

\bibitem{wernicke2006fanmod}
S.~Wernicke and F.~Rasche, ``Fanmod: a tool for fast network motif detection,''
  \emph{Bioinformatics}, vol.~22, no.~9, pp. 1152--1153, 2006.

\bibitem{de2015muxviz}
M.~De~Domenico, M.~A. Porter, and A.~Arenas, ``Muxviz: a tool for multilayer
  analysis and visualization of networks,'' \emph{Journal of Complex Networks},
  vol.~3, no.~2, pp. 159--176, 2015.

\bibitem{anchuri2018mining}
P.~Anchuri, M.~Berlingerio, and S.~Braghin, ``Mining relevant approximate
  subgraphs from multigraphs,'' Apr.~3 2018, uS Patent 9,934,327.

\bibitem{goyal2018graph}
P.~Goyal and E.~Ferrara, ``Graph embedding techniques, applications, and
  performance: A survey,'' \emph{Knowledge-Based Systems}, vol. 151, pp.
  78--94, 2018.

\bibitem{li2018multi}
J.~Li, C.~Chen, H.~Tong, and H.~Liu, ``Multi-layered network embedding,'' in
  \emph{Proceedings of the 2018 SIAM International Conference on Data
  Mining}.\hskip 1em plus 0.5em minus 0.4em\relax SIAM, 2018, pp. 684--692.

\bibitem{zhang2015multiple}
J.~Zhang and S.~Y. Philip, ``Multiple anonymized social networks alignment,''
  in \emph{2015 IEEE International Conference on Data Mining}.\hskip 1em plus
  0.5em minus 0.4em\relax IEEE, 2015, pp. 599--608.

\bibitem{hristova2015multilayer}
D.~Hristova, P.~Panzarasa, and C.~Mascolo, ``Multilayer brokerage in geo-social
  networks,'' in \emph{Ninth International AAAI Conference on Web and Social
  Media}, 2015.

\bibitem{rossi2015k}
L.~Rossi, M.~Musolesi, and A.~Torsello, ``On the k-anonymization of
  time-varying and multi-layer social graphs,'' in \emph{Ninth International
  AAAI Conference on Web and Social Media}, 2015.

\bibitem{vikatos2020marketing}
P.~Vikatos, P.~Gryllos, and C.~Makris, ``Marketing campaign targeting using
  bridge extraction in multiplex social network,'' \emph{Artificial
  Intelligence Review}, vol.~53, no.~1, pp. 703--724, 2020.

\bibitem{leskovec2010predicting}
J.~Leskovec, D.~Huttenlocher, and J.~Kleinberg, ``Predicting positive and
  negative links in online social networks,'' in \emph{WWW}.\hskip 1em plus
  0.5em minus 0.4em\relax ACM, 2010, pp. 641--650.

\bibitem{noel2011unfriending}
H.~Noel and B.~Nyhan, ``The “unfriending” problem: The consequences of
  homophily in friendship retention for causal estimates of social influence,''
  \emph{Social Networks}, vol.~33, no.~3, pp. 211--218, 2011.

\bibitem{magnani2013combinatorial}
M.~Magnani, B.~Micenkova, and L.~Rossi, ``Combinatorial analysis of multiple
  networks,'' \emph{arXiv preprint arXiv:1303.4986}, 2013.

\bibitem{coleman1957diffusion}
J.~Coleman, E.~Katz, and H.~Menzel, ``The diffusion of an innovation among
  physicians,'' \emph{Sociometry}, vol.~20, no.~4, pp. 253--270, 1957.

\bibitem{chen2006wiring}
B.~L. Chen, D.~H. Hall, and D.~B. Chklovskii, ``Wiring optimization can relate
  neuronal structure and function,'' \emph{PNAS}, vol. 103, no.~12, pp.
  4723--4728, 2006.

\bibitem{szell2010multirelational}
M.~Szell, R.~Lambiotte, and S.~Thurner, ``Multirelational organization of
  large-scale social networks in an online world,'' \emph{PNAS}, vol. 107,
  no.~31, pp. 13\,636--13\,641, 2010.

\bibitem{szell2010msd}
M.~Szell and S.~Thurner, ``Measuring social dynamics in a massive multiplayer
  online game,'' \emph{Social Networks}, vol.~32, pp. 313--329, 2010.

\bibitem{holme2002growing}
P.~Holme and B.~J. Kim, ``Growing scale-free networks with tunable
  clustering,'' \emph{Physical review E}, vol.~65, no.~2, p. 026107, 2002.

\bibitem{de2013mathematical}
M.~De~Domenico, A.~Sol{\'e}-Ribalta, E.~Cozzo, M.~Kivel{\"a}, Y.~Moreno, M.~A.
  Porter, S.~G{\'o}mez, and A.~Arenas, ``Mathematical formulation of multilayer
  networks,'' \emph{Phys Rev X}, vol.~3, no.~4, p. 041022, 2013.

\bibitem{hajibagheri2016holistic}
A.~Hajibagheri, G.~Sukthankar, and K.~Lakkaraju, ``A holistic approach for link
  prediction in multiplex networks,'' in \emph{International Conference on
  Social Informatics}.\hskip 1em plus 0.5em minus 0.4em\relax Springer, 2016,
  pp. 55--70.

\bibitem{rosvall2008maps}
M.~Rosvall and C.~T. Bergstrom, ``Maps of random walks on complex networks
  reveal community structure,'' \emph{PNAS}, vol. 105, no.~4, pp. 1118--1123,
  2008.

\bibitem{airoldi2008mixed}
E.~M. Airoldi, D.~M. Blei, S.~E. Fienberg, and E.~P. Xing, ``Mixed membership
  stochastic blockmodels,'' \emph{Journal of machine learning research},
  vol.~9, no. Sep, pp. 1981--2014, 2008.

\bibitem{roxana2019edge}
A.~Roxana~Pamfil, S.~D. Howison, and M.~A. Porter, ``Edge correlations in
  multilayer networks,'' \emph{arXiv:1908.03875}, 2019.

\bibitem{perozzi2014deepwalk}
B.~Perozzi, R.~Al-Rfou, and S.~Skiena, ``Deepwalk: Online learning of social
  representations,'' in \emph{Proceedings of the 20th ACM SIGKDD international
  conference on Knowledge discovery and data mining}, 2014, pp. 701--710.

\bibitem{grover2016node2vec}
A.~Grover and J.~Leskovec, ``node2vec: Scalable feature learning for
  networks,'' in \emph{Proceedings of the 22nd ACM SIGKDD international
  conference on Knowledge discovery and data mining}, 2016, pp. 855--864.

\bibitem{zhou2009predicting}
T.~Zhou, L.~L{\"u}, and Y.-C. Zhang, ``Predicting missing links via local
  information,'' \emph{The European Physical Journal B}, vol.~71, no.~4, pp.
  623--630, 2009.

\bibitem{adamic2003friends}
L.~A. Adamic and E.~Adar, ``Friends and neighbors on the web,'' \emph{Social
  networks}, vol.~25, no.~3, pp. 211--230, 2003.

\bibitem{heider1958psychology}
F.~Heider, ``The psychology of interpersonal relations,'' 1958.

\bibitem{antal2005dynamics}
T.~Antal, P.~L. Krapivsky, and S.~Redner, ``Dynamics of social balance on
  networks,'' \emph{Phys Rev E}, vol.~72, no.~3, p. 036121, 2005.

\bibitem{kirkley2019balance}
A.~Kirkley, G.~T. Cantwell, and M.~E. Newman, ``Balance in signed networks,''
  \emph{Physical Review E}, vol.~99, no.~1, p. 012320, 2019.

\bibitem{harary1959measurement}
F.~Harary, ``On the measurement of structural balance,'' \emph{Behavioral
  Science}, vol.~4, no.~4, pp. 316--323, 1959.

\bibitem{aref2019balance}
S.~Aref and M.~C. Wilson, ``Balance and frustration in signed networks,''
  \emph{Complex Networks}, vol.~7, no.~2, pp. 163--189, 2019.

\bibitem{aref2016exact}
S.~Aref, A.~J. Mason, and M.~C. Wilson, ``An exact method for computing the
  frustration index in signed networks using binary programming,'' \emph{arXiv
  preprint arXiv:1611.09030}, p.~55, 2016.

\bibitem{laifa2015overview}
M.~Laifa, S.~Akrouf, and R.~Maamri, ``An overview of forgiveness in the digital
  environment,'' in \emph{Proceedings of the International Conference on
  Intelligent Information Processing, Security and Advanced Communication},
  2015, pp. 1--5.

\end{thebibliography}

\vspace{1em}

\begin{footnotesize}

\begin{wrapfigure}{l}{.2\columnwidth}
\includegraphics[width=.2\columnwidth]{{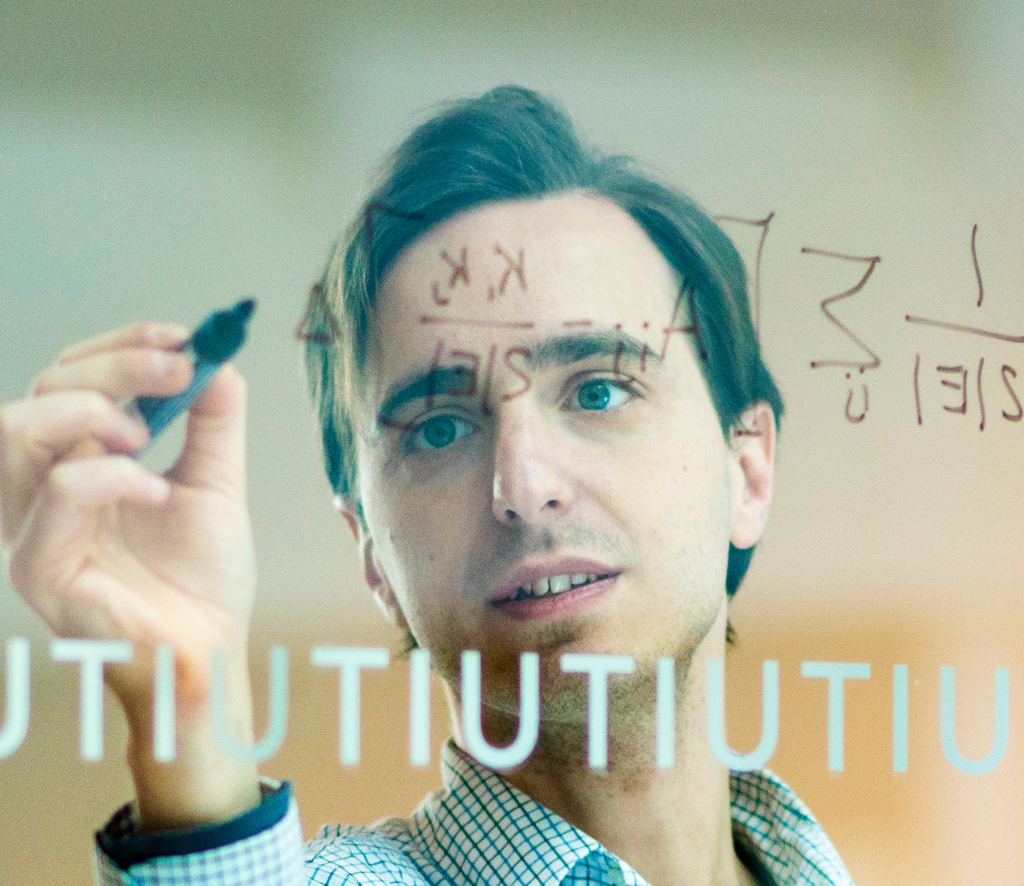}}
\end{wrapfigure}

Michele Coscia is an associate prof at IT University of Copenhagen. He works on algorithms for the analysis of complex networks, and on applying the extracted knowledge to a variety of problems. His background is in Digital Humanities. He holds a PhD in Computer Science, obtained in 2012 at the University of Pisa. In the past, he visited CCNR at Northeastern University, and worked for 6 years at the Center for International Development, Harvard University. 

\vspace{.75em}

\begin{wrapfigure}{l}{.2\columnwidth}
\includegraphics[width=.2\columnwidth]{{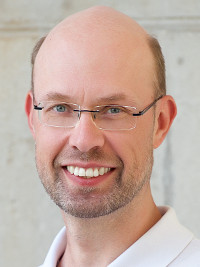}}
\end{wrapfigure}

Christian Borgelt received his M.Sc. in computer science from the Technical University Carolo-Wilhelmina of Braunschweig, Germany, in 1995. After spending a year at the Daimler-Benz Research Center Ulm, he became a Ph.D. student at the University of Magdeburg, Germany, and received his Ph.D. in 2000. In 2006 he received the venia legendi for computer science, again from the University of Magdeburg, Germany. From April 2006 to May 2016 he was the principal researcher of the Intelligent Data Analysis and Graphical Models Research Unit of the European Centre for Soft Computing, Mieres, Spain. After a year as a freelance IT consultant, he became a substitute professor at the University of Konstanz, Germany, from April 2017 to September 2018. Since October 2018 he is Professor for Data Science at the Paris-Lodron-University of Salzburg in Austria.

\vspace{.75em}

\begin{wrapfigure}{l}{.2\columnwidth}
\includegraphics[width=.2\columnwidth]{{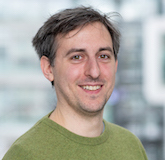}}
\end{wrapfigure}

Michael Szell is associate professor in Data Science at IT University of Copenhagen, and external researcher at ISI Foundation and at the Complexity Science Hub Vienna. His research quantifies the patterns behind interlinked human behavior and human-built structures through mining large-scale data sets. He follows an anti-disciplinary approach using methods from data science and network science. Michael's current focus is on sustainable urban mobility and urban data science. He has developed interactive data visualization platforms such as What the Street!?, and the award-winning MMO game Pardus.

\end{footnotesize}

\end{document}